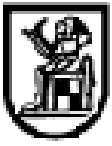 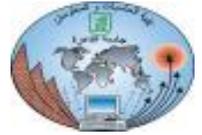

Cairo University

Faculty of Computers & Information

Department of Information Systems

# Knowledge Discovery In GIS Data

Submitted By

**Ayman Taha Awad-Allah Mohamed Farahat**

a.taha@fci-cu.edu.eg

## Under supervision of

**Prof. Osman Mohamed Hegazi**

Professor, Faculty of Computers & Information, Information Systems Dept., Cairo University,
o.hegazi@fci-cu.edu.eg

**Dr. Hoda Mohamed Onsi**

Associate Professor, Faculty of Computers & Information, Information Technology Dept., Cairo University,
h.onsi@fci-cu.edu.eg

**Dr. Mohamed Nour El_dein**

Assistant Professor, Faculty of Computers & Information, Information Systems Dept, Cairo University
m_nour@aucegypt.edu

A thesis submitted to the faculty of computers and information Cairo University in of the degree of Master of Science in Information Systems

December 2005

# Abstract

Intelligent geographic information system (IGIS) is one of the promising topics in GIS field. It aims at making GIS tools more sensitive for large volumes of data stored inside GIS systems by integrating GIS with other computer sciences such as Expert system (ES) Data Warehouse (DW), Decision Support System (DSS), or Knowledge Discovery Database (KDD). One of the main branches of IGIS is the Geographic Knowledge Discovery (GKD) which tries to discover the implicit knowledge in the spatial databases.

The main difference between traditional KDD techniques and GKD techniques is hidden in the nature of spatial datasets. In other words in the traditional dataset the values of each object are supposed to be independent from other objects in the same dataset, whereas the spatial dataset tends to be highly correlated according to the first law of geography. The spatial outlier detection is one of the most popular spatial data mining techniques which is used to detect spatial objects whose non-spatial attributes values are extremely different from those of their neighboring objects. Analyzing the behavior of these objects may produce an interesting knowledge, which has an effective role in the decision-making process.

In this thesis, a new definition for the spatial neighborhood relationship by is proposed considering the weights of the most effective parameters of neighboring objects in a given spatial dataset. The spatial parameters taken into our consideration are; distance, cost, and number of direct connections between neighboring objects. A new model to detect spatial outliers is also presented based on the new definition of the spatial neighborhood relationship. This model is adapted to be applied to polygonal objects. The proposed model is applied to an existing project for supporting literacy in Fayoum governorate in Arab Republic of Egypt (ARE).

# LIST OF TABLES

# LIST OF Figures

# Table of Contents

# Chapter 1

# Spatial Data Bases (SDB)

1.1 Introduction

1.2 GIS importance

1.3 Spatial data models in GIS

1.4 Mapping concepts, features & properties

1.5 GIS data types

1.6 Conclusion

## 1.1 Introduction

A spatial database is a collection of spatially referenced data that acts as a model of reality [Shekhar et al., 2003b]. A spatial database can be defined as " a database system that offers spatial data types in its data model and query language and supports spatial data types in its implementation, providing at least spatial indexing and spatial join methods"[Guting, 1994].

Spatial databases can be used in many fields such as [Shekhar et al., 1999]:-

- **Physical world** (geography, urban planning, astronomy,…)
- **Parts of living organisms** (anatomy of the human body,…)
- **Engineering design** (integrated circuits, automobile design,…)
- **Multimedia database** (image and video processing,…)

Geographic Information Systems (GIS) is considered one of the most important applications that depends on spatial databases for storing its georeferenced data objects. GIS is defined as "a system for capturing, storing, checking, integrating, manipulating, analysis, and displaying data which are spatially referenced to the earth". [Hughes 1994].

We should make a distinction between the two words; spatial and geographic. Spatial concerns any phenomena where the data objects can be embedded within some formal space that generates implicit relationships among the objects such as multimedia images and astronomy. But geographic refers to the specific case where the data objects are georeferenced and the embedding space relates to locations on or near the earth's surface [Miller 2003].

## 1.2 GIS importance

GIS is suitable for many kinds of applications and can be used to solve many problems such as [NCGIA 1990]:-

- Geographical and spatial analysis which is needed to represent the regional-geographic and spatially problems.
- Understanding and managing location based applications.

- Analyzing geographical phenomena by integrating spatial and non- spatial data within a single system.
- Manipulating and displaying geographical knowledge in new and exciting ways.

GIS is considered a convergence of technological fields and traditional disciplines such as geography, cartography, remote sensing, photogrammetry, surveying, statistics, computer science , and mathematics.

## 1.3 Spatial data models in GIS

There are two different data models in GIS to represent geometrical data: raster and vector. The type of data models affects both data storage volume and processing speed.

### 1.3.1 Raster data model

A raster data set is a regular grid of cells divided into rows and columns where data values for a given parameter are stored in each cell as depicted in (Figure 1-1). The spatial resolution of the raster data set depends on the size of the cell. So the spatial resolution should be accurately determined before data collection phase depending on the desired outcomes [NCGIA 1990].

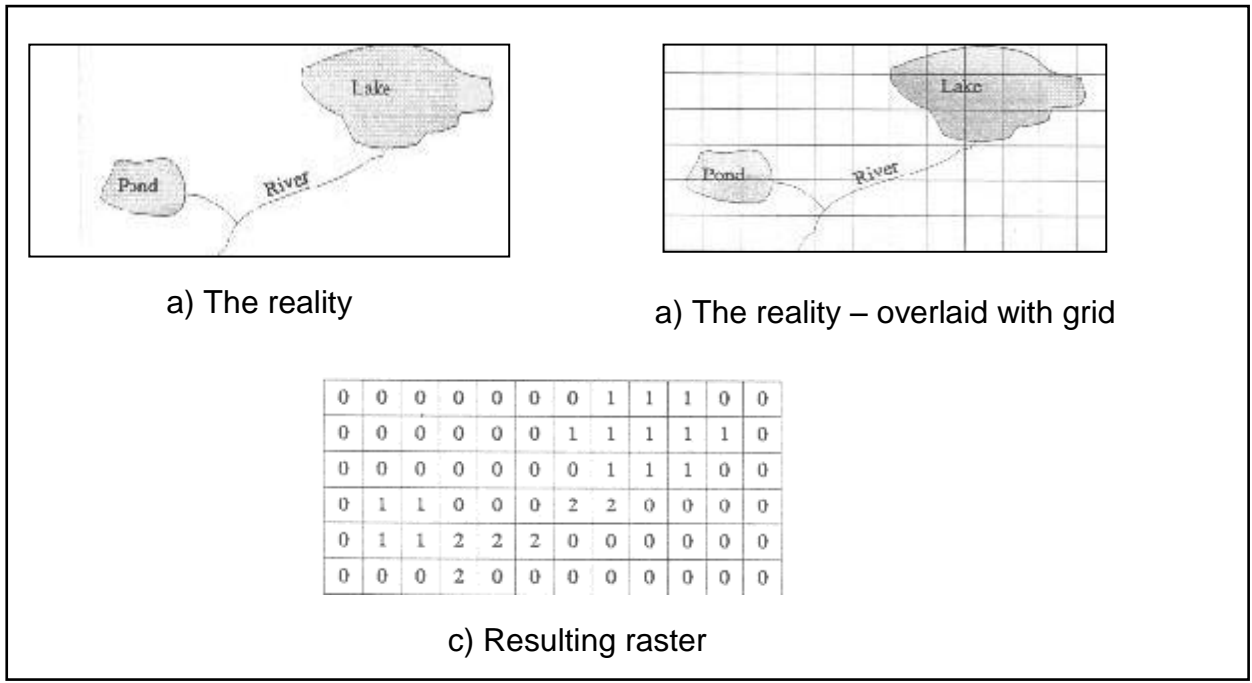

a) The reality

a) The reality – overlaid with grid

c) Resulting raster

Figure 1-1: Raster data model structure

[Star et al, 1990]

Raster data acquisition methods are easier than vector data acquisition methods because there are many simple ways to get raster data such as digital images, scanning maps and remote sensing images. Raster data model is appropriate for applications where a continuous space is used such as soil type, rainfall, temperature, or elevation. On contrary, raster data model is inappropriate for applications where discrete boundaries must be known, such as parcel management.

## 1.3.2 Vector data model

Instead of storing data about each cell in the map in raster data model, vector data model stores data about interested features only, so the size of a map in raster model is larger than the same map if it stored in vector model. Vector model classify the objects into 3 categories; points, arcs, and polygons. Points are considered fundamental primitives in vector data model. Lines are created by connecting points with straight lines or arc of circle and so polygons are defined by sets of arcs as depicted in (Figure 1-2).

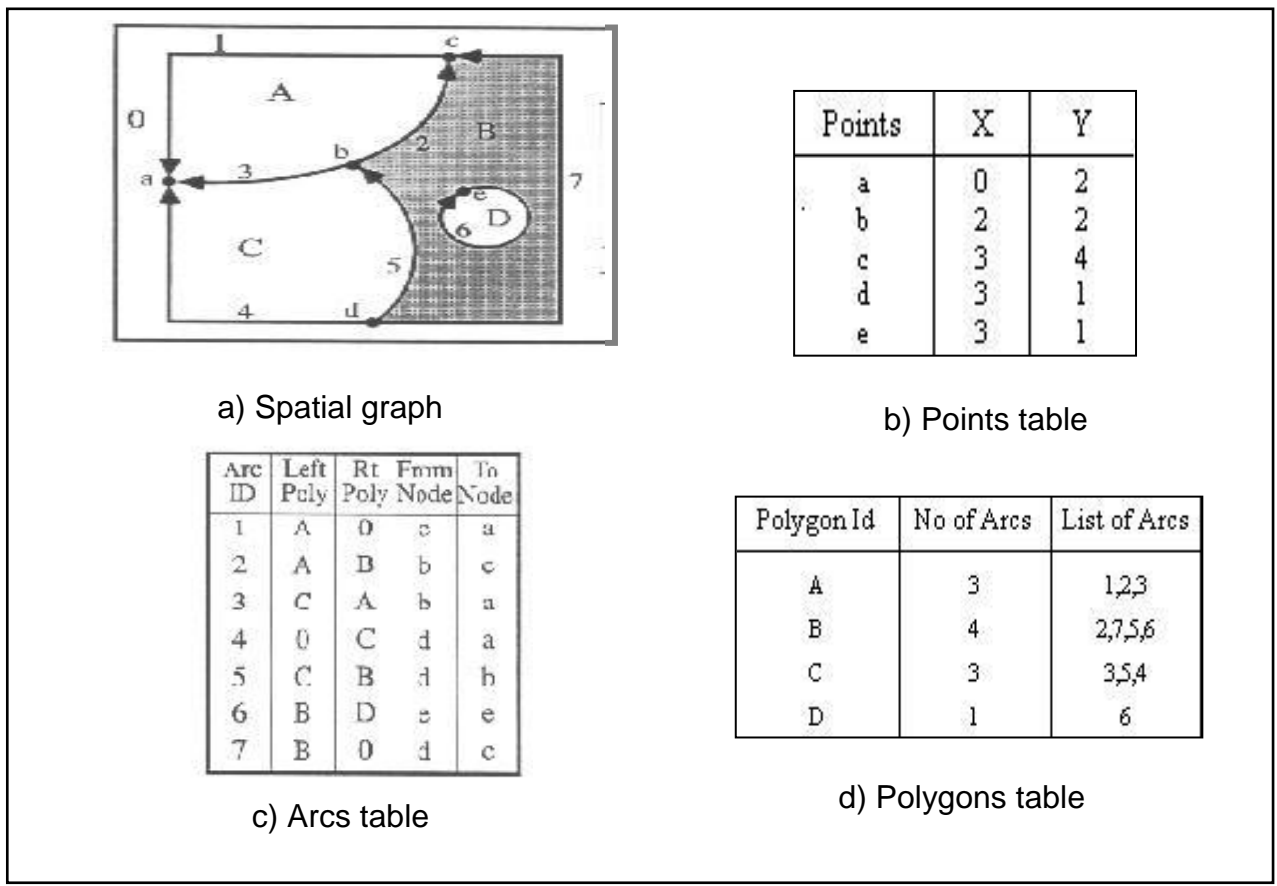

Figure 1-2: Vector data model structure

[NCGIA 1990]

## 1.4 Mapping concepts, features and properties

GIS abstracts reality to a set of maps. A map represents geographic features or other spatial phenomena by graphically conveying information about locations. The basic objectives of the map are to provide [NCGIA 1990]:-

- Descriptions of geographic phenomenon.
- Spatial and non spatial information.
- Map features such as point, line, and polygon.

### 1.4.1 Map Scale

Map scale is the ratio between distances on the map and the corresponding distances in the reality. If a map scale is 1:10,000, then 1 cm on the map equals 10,000 cm or 100 m on the Earth's surface. There is confusion between the use of the terms "small scale" and "large scale"; a large scale map shows great detail [Davis 1996].

### 1.4.2 Map features

A GIS maps objects in reality to a set map features. Map features could be classified into three categories points, lines and polygons.

**Point feature**

A point feature represents a single location such as houses, customers, and trees as depicted in (Figure 1-3). A point is represented as a pair of coordinates (X,Y).

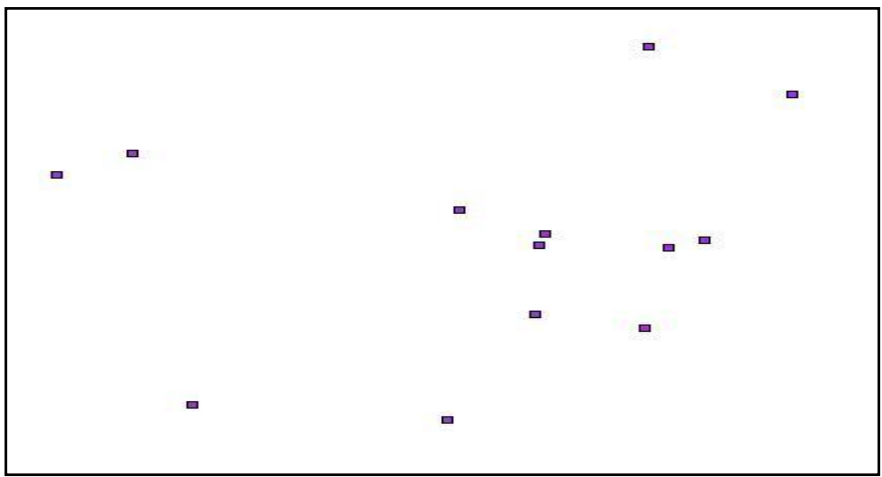

12
Figure 1-3: Samples of point features

[ESRI 2003]

## Line feature

A line feature is a set of connected, ordered coordinates representing the linear shape of a map object such as roads, streets, rivers and administrative boundary between villages as depicted in (Figure 1-4).

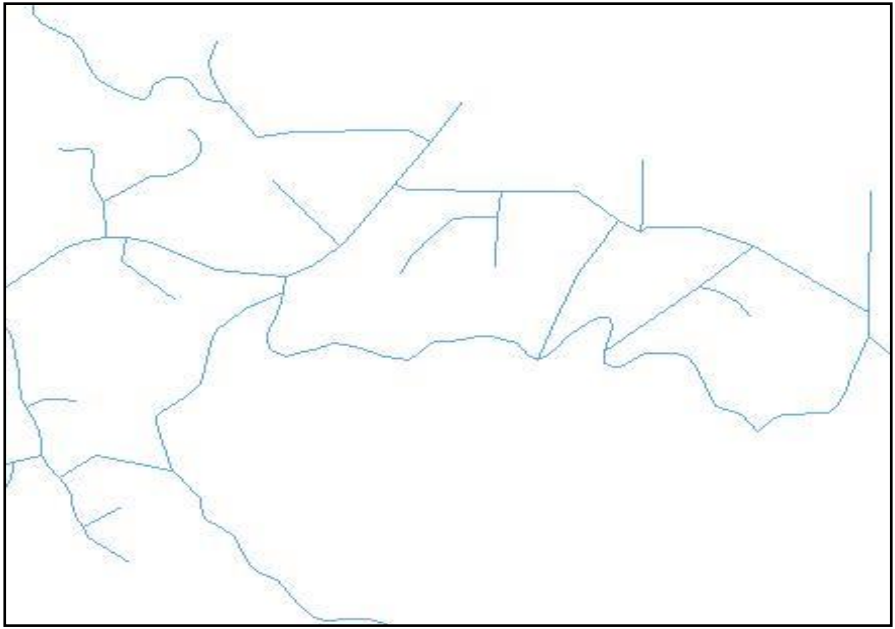

Figure1-4: Samples of line features

[ESRI 2003]

## Polygon feature

A polygon feature is a closed figure whose boundary encloses a homogeneous area, such as states, lakes, and villages as depicted in (Figure 1-5).

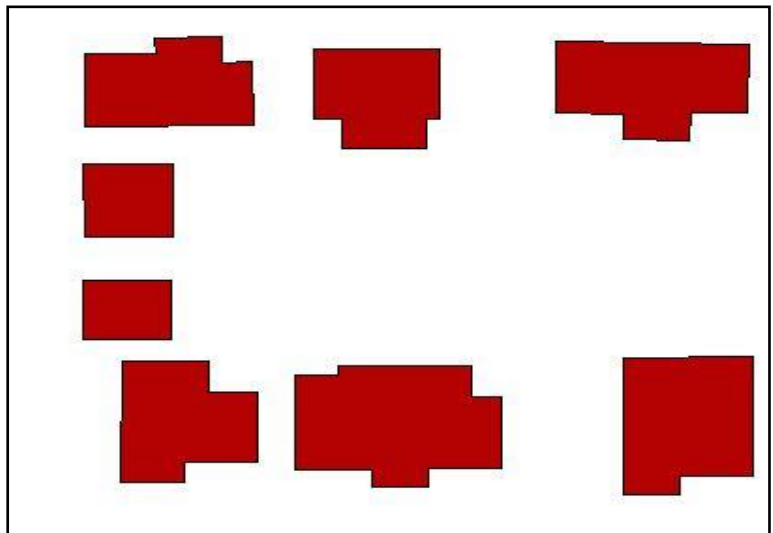

13
Figure 1-5: Samples of polygon features

[ESRI 2003]

### 1.4.3 Transforming between map features

Transformation between map features is a common process in GIS. There are two types of transformations; increasing dimensionality and decreasing dimensionality.

Buffering process is an example of increasing dimensionality which is used to transform points to polygons by drawing buffers around this point. Buffering process can also be used to transform lines to polygons by drawing buffers around lines. On the other hand, centroid process is an example of decreasing dimensionality is which is used to transform polygons to points. These points always are the centers of these polygons [Harvard 2005].

Buffering and centroid processes may cause changing in the map scale. In this thesis, centroid process is used to calculate distance between two neighboring villages.

### 1.4.4 GIS concepts

The power of a GIS over paper maps is the ability to select the needed information to be seen according to the goal wanted to be achieved. GIS has two main concepts; features have attributes associated with them and features are separated into layers [ESRI 2003].

- **Features have attributes associated with them:**

Each map features should have its own record in database. Information about this map feature is stored in this record. Each map feature should have object_Id which is used to tie between this map feature and its information stored in database.

- **Features are separated into layers**

GIS divides a large map into manageable pieces called layers. For example, all roads could be on one layer and all customers on another layer. GIS should have the ability to relate and join the layers in the same application as depicted in (Figure 1-6).

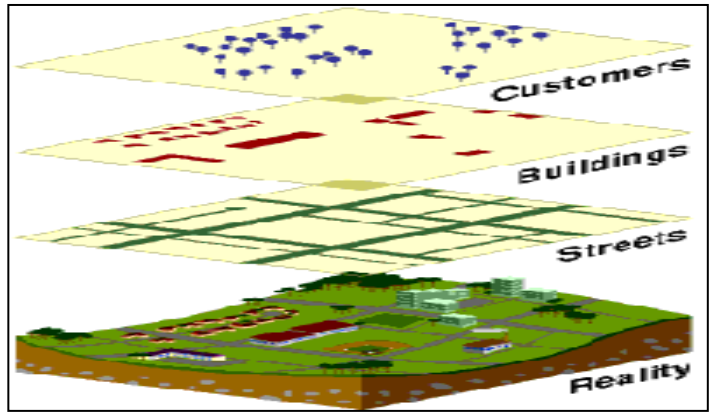

Figure 1-6: Layers representing the real world

[ESRI 2003]

## 1.5 GIS data types

GIS combines information about a place to provide a better understanding of that place. These pieces of information are gathered from many sources such as photos, maps, cad files, relational databases, and spread sheets as depicted in (Figure 1-7).

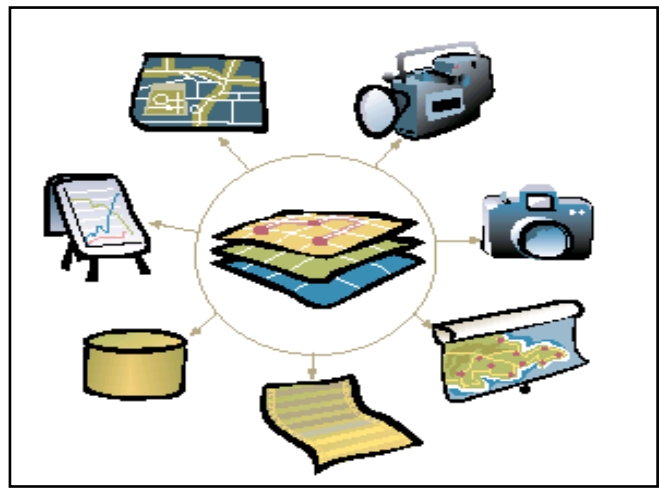

Figure 1-7: Data sources in GIS

[ESRI 2003]

Data in a GIS comes in three basic forms
- Geometrical

- Topological
- Attribute

### 1.5.1 Geometrical data

The geometrical data in a digital map provides the location and the form of each map feature. For example, the geometrical data in a road map are the location and the shape of each road on the map.

In a vector map, a feature's location is normally expressed in a 2D GIS models as sets of X and Y pairs or X, Y and Z triples in a 3D GIS models. Most vector geometrical data systems support three fundamental geometric objects; point, line and polygon. But some systems also support more complex entities, such as regions, circles, ellipses, arcs, and curves [NCGIA 1990]

### 1.5.2 Topological data

It describes the most useful relationships among spatial features. It is defined as "a mathematical procedure for explicitly defining spatial relationships among spatial features" [Lakhan 1996]. The topological data logically determines exactly how and where points and lines are connected in a map by means of nodes. Topological data provides additional intelligence for manipulating, analyzing, and using the information stored in a database [Burrough 1986]. These relationships can be classified into six categories according to the data types of the features participate in this relationship [NCGIA 1990]:-

1. **Point-point relationships**

    - **"IS WITHIN"**, e.g. find all of the customer points within 1 km of this retail store point.
    - **"IS NEAREST TO"**, e.g. find the nearest hospital to accident location.

2. **Point-line relationships**
    - **"ENDS AT"**, e.g. find the intersection at the end of this street.
    - **"IS NEAREST TO"**, e.g. find the road nearest to this aircraft crash site.

3. **Point-polygon relationships**
   - **"IS CONTAINED IN"**, e.g. find all customers located in this ZIP code boundary.
   - **"CAN BE SEEN FROM"**, e.g. determine if any of this lake can be seen from this viewpoint.

4. **Line-line relationships**
   - **"CROSSES"**, e.g. determine if this road crosses this river.
   - **"COMES WITHIN"**, e.g. find all of the roads which come within 1 km of this railroad.

5. **Line-polygon relationships**
   - **"CROSSES"**, e.g. find all of the soil types crossed by this railroad.
   - **"BORDERS "**, e.g. find out if this road forms part of the boundary of this airfield.

6. **Polygon-polygon relationships**
   - **"OVERLAPS"**, e.g. identify all overlaps between types of soil on this map and types of land use on this other map.
   - **"IS NEAREST TO "**, e.g. find the nearest lake to this forest fire.
   - **"IS ADJACENT TO"**, e.g. find out if these two polygons share a common boundary.

**Relationships between spatial features may be divided into three types [Keating et al., 1987]: -**

- Relationships which are used to construct complex objects from simple primitives such as the relationship between a line and a chain of ordered points which define it.

- Relationships which can be computed from the coordinates of the objects such as the "crosses" relationship between two lines and the "is contained in" relationship checks if a given point inside a certain polygon or not.

- Relationships which cannot be computed from coordinates - these must be coded in the database during the input phase such as "an overpass" relationship that shows a bridge which overpasses a given highway.

### 1.5.3 Attribute data

It is the stored data which describes map features. For example, an attribute associated with a road might be its name or number of lanes. Attribute data is often stored in database files which is kept separately from the graphic portion of the map. GIS tools should maintain internal links tying each spatial map feature to its attribute information. Each map feature has an attached key value stored with it. This key value identifies the specific database record that contains the feature's attribute information [Lakhan 1996].

Topological and attributes data are always associated with the vector maps, they are seldom associated with raster images, because the attribute data stores description of spatial objects and topological data stores the relationships between them. But the concept of objects is not the nature of raster maps which model the world as composed of pixels [Davis 1996].

## 1.6 Conclusion

GIS is considered the most famous application that uses spatial databases to store, capture, integrate, manipulate, analyze, and display data which are spatially referenced to the earth. GIS is one of early information systems that benefits from the technological revolution in the second part in previous century. Data in spatial domain are categorized into three classes geometrical, topological and attribute data; geometrical and topological are the main features of spatial databases that differs them from other traditional databases systems.

There are two models to store geometrical data; raster data model and vector data model. Raster data are easy to get but topological in raster data are more complex than vector data. In addition, analysis methods are also very poor in raster data because they depend on the amount of topological data stored in spatial database. It makes so knowledge based applications in spatial domain are preferable to use vector data.

In past, large amount of data stored in spatial databases and the complexity of geometric operations make the scope of GIS applications to move away from powerful analysis methods; so the main objectives of GIS was visualization. But recently, the great improvement in hardware, analysis methods and mathematical operations made geographic researchers to direct their research towards analysis of spatial databases.

In this thesis we focus on vector data model and how to use topological data to discover knowledge from these data. But our model can be adapted to be beneficially applied into raster model. We also use centroid process to transform polygons to points to calculate distance between two polygons. The next chapter presents knowledge discovery process, then we will project this process into spatial domain.

# Chapter 2

# Knowledge discovery process

2.1 Introduction

2.2 Knowledge discovery motivation

2.3 The knowledge discovery process

2.4 Data warehouses

2.5 Data mining

2.6 Conclusion

## 2.1 Introduction

The explosive growth in amount of data stored in databases, document files, web pages, and any other digital formats has made the manual and traditional methods of analyzing and extracting information which are used in the business management process to be quite difficult. Scientists in data analysis and databases are directed to automatic methods for extracting hidden information in large amount of data. Such extracted information is rather useful in decision making process [Zhang et al., 2002].

Knowledge Discovery in Databases (KDD) is a process of extracting hidden knowledge from large amount of data. Some people treat Data Mining (DM) as a synonym to KDD but the last is considered a high level process for obtaining facts through data mining then distilling this information into knowledge and beliefs. So data mining is considered as part of KDD process. [Han et al., 2001], [Miller al., 2001] and [Miller 2003].

## 2.2 Knowledge discovery motivations

The classical approach of data analysis is based on one or more analysts. Data analysts should be familiar to the data and should serve as an interface between the data and the users. In fact, as data volumes grow rapidly, this method becomes completely impractical in many applications [Fayyad et al., 1996].

**The motivations of Knowledge Discovery in Databases (KDD) are** [Williams et al., 1998]**:-**

1. Decisions must be made with maximum knowledge.
2. Decisions must be made rapidly.
3. The growing customers demands.
4. Huge databases.
5. Databases are growing at an unexpected rate.
6. Cheaper and faster hardware.

## 2.3 The knowledge discovery process

The KDD is interactive and iterative process, involving numerous steps; many decisions are made by the users. (Figure 2-1) shows the KDD process contains; background, data pre-processing and data warehousing, data preparation, data mining, knowledge evaluation and deployment [Miller, 2003].

### 2.3.1 Background:

Developers should have good background knowledge of the applications domain. This step may contain the following tasks:-

- **Gathering data:** Collecting data that will be used in analysis. This may include the acquisition of external data from public databases.

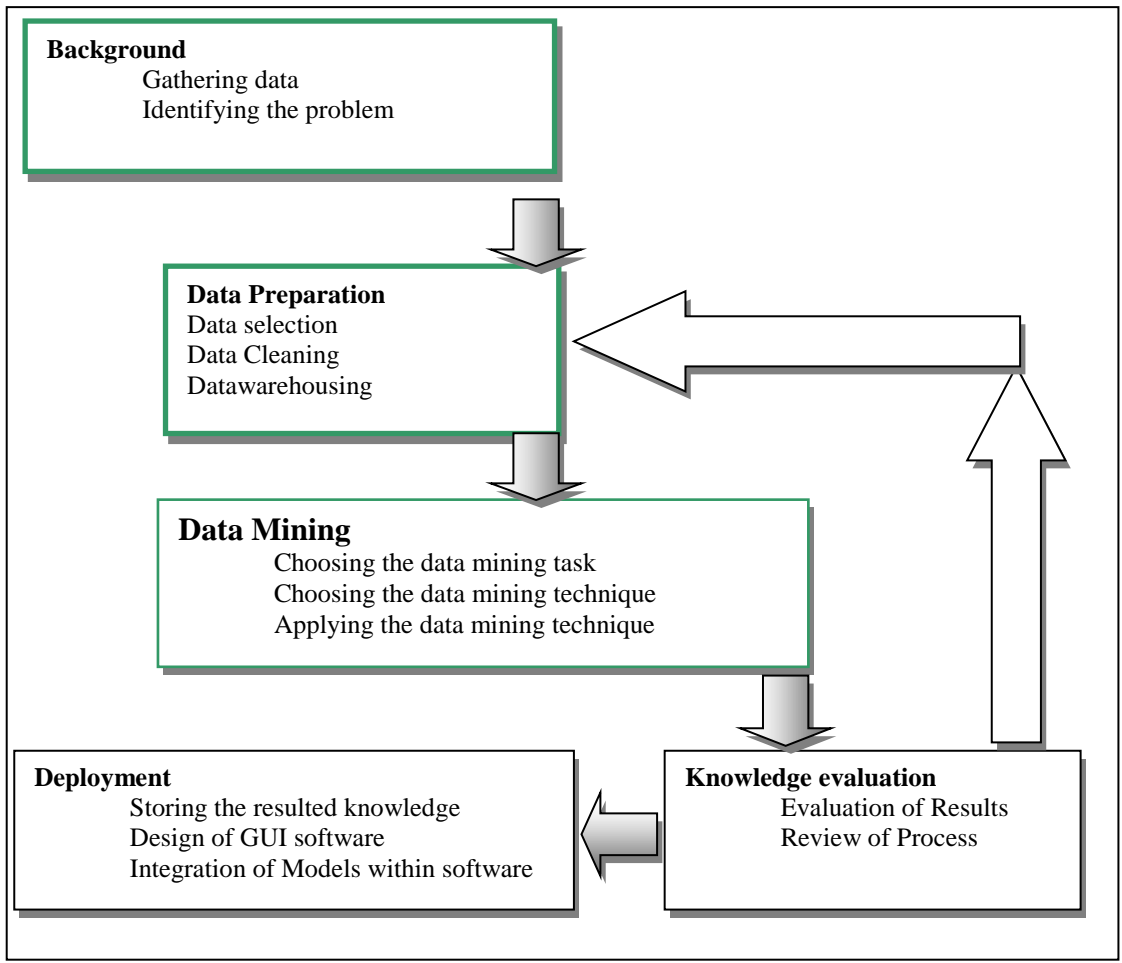

Figure 2-1: Block diagram of knowledge discovery framework

- **Identifying the problem**: Before starting knowledge discovery process must make a clear statement of the objectives depending on the problem statement. An effective statement of the problem will also include a way of measuring the resulted knowledge [Miller, 2003].

### 2.3.2. Data Preparation:-
- **Data selection**: Determining a subset of data for focusing the search for patterns of interest.
- **Data Cleaning**: Cleaning data by removing noise and inconsistent data.
- **Data warehousing**: Including integrating, transformations, projections and aggregations to find useful representations for the data.

### 2.3.3 Data mining: Searching for hidden, implicit, and significant patterns or trends that would be useful knowledge.

### 2.3.4 Knowledge Evaluation:
- Evaluating the mined patterns, often through visualization.
- Reviewing the knowledge discovered process,

### 2.3.5 Deployment [Qi et al., 2003]:
- Storing the resulted knowledge either by incorporating the knowledge into a computational system (such as a knowledge-based database) or through documenting and reporting the knowledge to interested reports.
- Designing user friendly GUI to be used by decision makers

## 2.4 Data warehouses

DWs are considered one of the hottest topics in the area of information systems. Data warehouses are important preliminary step in KDD process especially when having many data sources. DW is a "collection of subject-oriented, integrated, non-volatile and time-variant data to support decision making process"[Inmon 1996]. DW is used as a large storage area for data which are collected from different data sources. These data sources may be database, document file, xml, or html. The main objective of making DWs is to analyze these data to support the mangers to make decisions [Zimányi 2003].

In addition to the DW, the data mart term is used, data mart may be a part of the data warehouse. The definition of a data mart is expressed identically to the definition of the DW except that all the DW's features are applied to small part of business that can be department (e.g., marketing department), a certain activity (e.g., analysis of product promotion) or increase the level of summarization of data in the DW [Han et al., 2001].

## 2.4.1 Data warehousing architecture

The goal of making DW is to transform a huge, inconsistent, and separated data into integrated repository and to use this repository to produce information. Such information is helpful in decision making process. As shown in (Figure 2-2), the conversion of data into useful information could be done using a variety of different tools and applications [Zimányi 2003]:-

- **Report and applications**: The developers could create applications or reports directly from DW according to the user requirements using the traditional tools for that process.
- **OLAP:** Provides dynamic form to answer online user's query, the user can prepare ad-hoc queries applying one of the following capabilities:-
    - **-Drill-down**: allows automatic transformation of summarized (aggregated) measures into a more detailed form, for example transforming numbers of illiterates in city to or numbers of illiterates in its villages.
    - **-Roll-up:** In contrast drill-down allows transforming detailed measures into summarized data.
    - **-Slicing-and-dicing:** allows visualizing only the information satisfying specified conditions. For example, number of population in the Cairo governorate during the year of 2000 [Zimányi 2003].

- **Statistics:** traditional tools that allow the user to understand data characteristics using distribution, correlation, or any other statistical reasoning analysis.

- **Data mining:** Intelligent techniques that allow user to discover hidden knowledge in large or rich data base [Zimányi 2003].

## 2.4.2 General characteristics of data warehouse

DW should have the following components [Kimbal 1996]:

- **Measures;** are a set of attributes representing the core of analysis such as quantity sold, sales, cost, and number of illiterates [Inmon 1996].

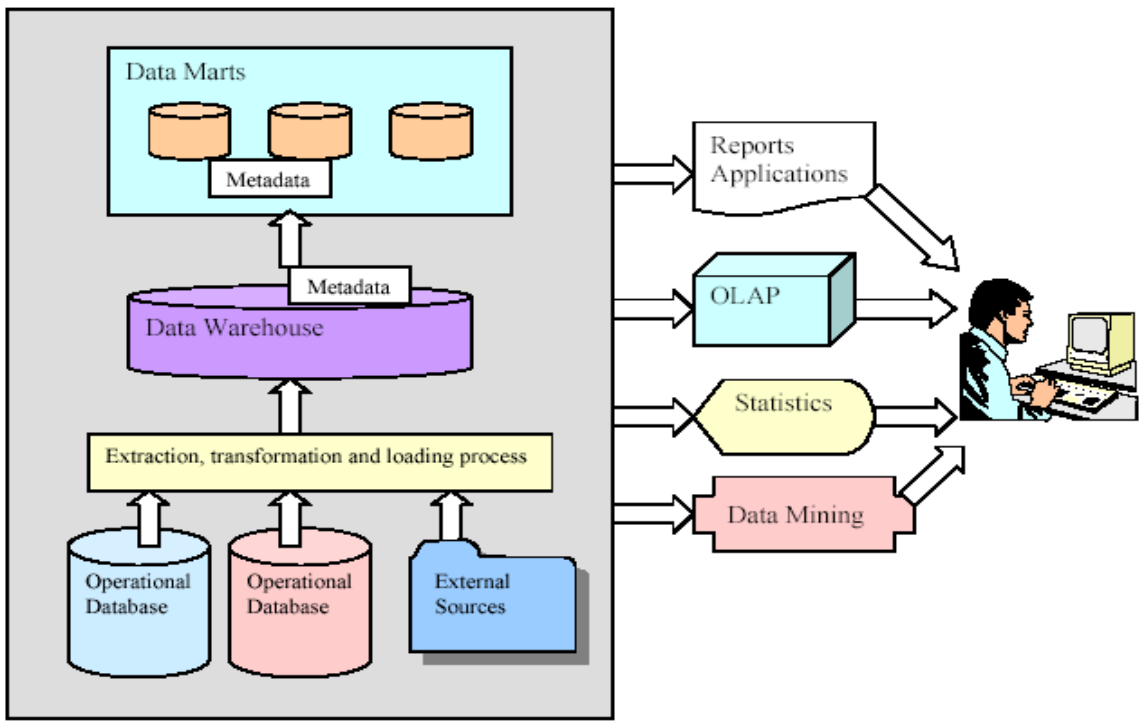

Figure 2-2: Data warehousing architecture

[Zimányi 2003]

- **Fact table;** represents the subject-orientation and the focus of analysis. It typically contains measures. In addition it contains the all keys of the dimension tables to allow the fact measures to be analyzed from different perspectives [Pedersen et al., 2000] as depicted in (Figure 2-3).

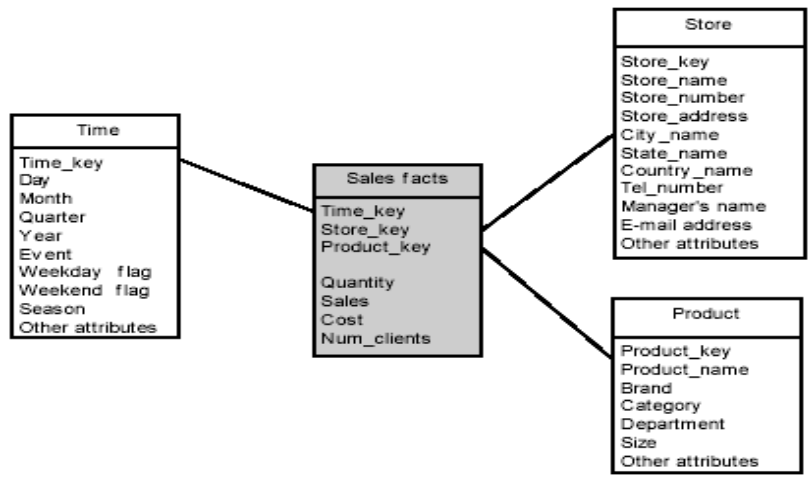

Figure 2-3: Sales DW schema

- **Dimension tables;** are object tables that include attributes a [Zimányi 2003] explore the measures from different analysis perspectives [Kimball et al., 98]. For example, the number_of_population in a population DW can be seen during different periods of time (time dimension) or in different parts of the country (location dimension). These dimensions are connected to the fact table in the relational model through foreign keys [Golfarelli et al., 99] and [Theodoratos et al., 99]

- **Hierarchies;** the dimension attributes may form a hierarchy, such as City – State – Country in Store dimension as depicted in (Figure 2-4). Allowing the user to see detailed as well as aggregated data. The importance of having well defined hierarchies in a DW environment is due to the fact that, the decision-maker usually starts from a general view of data and then goes to the detailed explorations. Hierarchies are usually presented in logical level using a flat table called star shema or using a normalized structure called snowflake scheme [Eder et al., 2002].

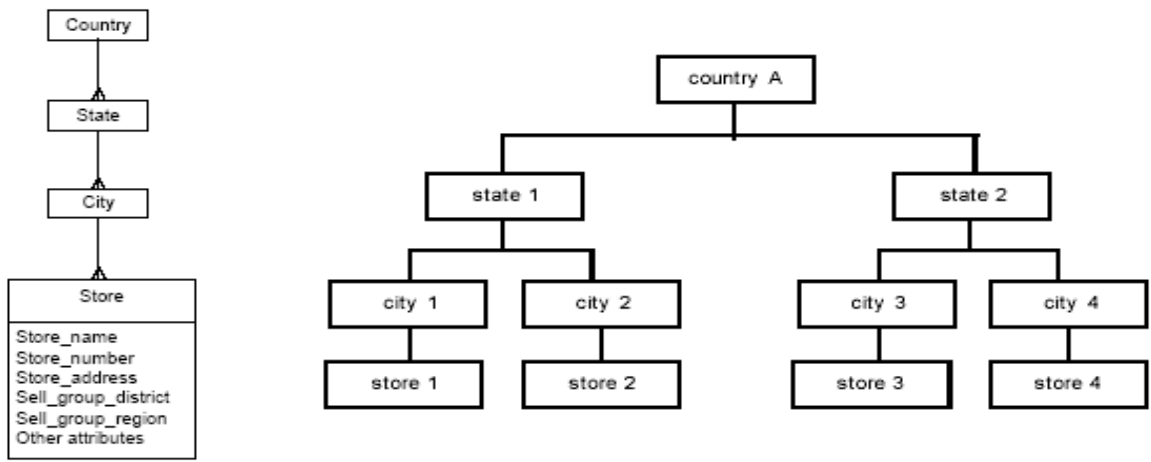

Figure 2-4: Store dimension hierarchy

[Zimányi 2003]

## 2.5 Data mining

Most organizations have accumulated a great deal of data, whereas what they really want is information. How can they learn from the data about how to satisfy their customers? How to allocate their resources most efficiently? And how can one draw meaningful conclusions about the forest? Data mining is used to address these concerns [Fayyad et al., 1996].

Data mining uses sophisticated statistical analysis and modelling techniques to discover hidden patterns in data stores. Such patterns are usually missed by ordinary analysis methods. The goal of the entire KDD process is to make patterns understandable to humans in order to give a better interpretation of data [Williams et al., 1998].The of data mining is to find these patterns and relationships among them by building models. Data mining may be considered the gathering point of other computer science branches such as machine learning, databases, visualization, statistics, and parallel algorithms as depicted in (Figure 2-5).

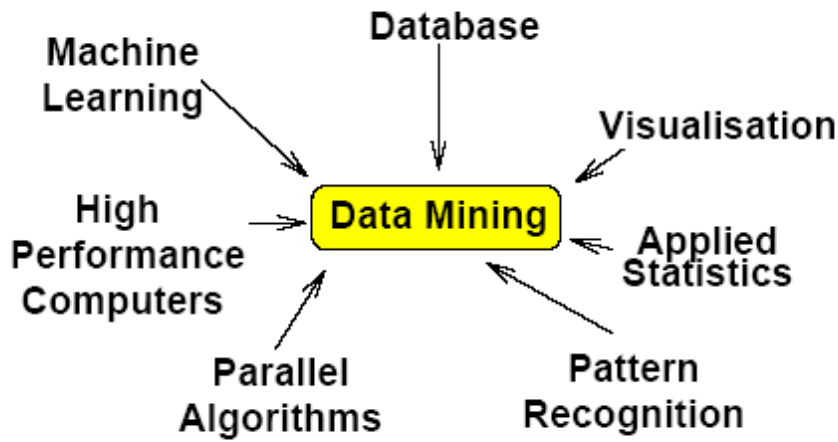

Figure 2-5: Relationship among data mining and other computer sciences

[Williams et al., 1998]

### 2.5.1 Steps of data mining process

The process of data mining has been done through three sub steps [Han et al 2001]:-

- Choosing the data mining task**;** this involves selecting the generic type of pattern through data mining; this is the patterns for expressing facts in the database. Generic pattern types include classes, association rules, clusters, outlier detection.
- Choosing the data mining technique for discovering patterns of the generic type selected in the previous step. Data mining algorithms are often heuristics due to scalability requirements, there are typically several techniques available for a given pattern type.
- Applying the data mining technique to search for interesting patterns.

### 2.5.2 Data mining techniques

Most data mining techniques are heuristics tailored to discover patterns of a generic type. Generic patterns include classes, associations, rules, clusters and outlier detection. Since these techniques are heuristics, there is no single optimal algorithm for discovering patterns of a given type. Different techniques highlight different aspects of the information space. [Han et al 2001].

**2.5.2.1 Classification**

Classification is a well known data mining technique and it has been studied in machine learning community for a long time. Classification involves mapping data into specified classes or meaningful categories based on common properties among a set of objects in a database [Sousa et al., 1998].The cardinality of these classes is much less than the number of data objects. After the construction of the classification model, it is used to classify new cases that are going to be inserted in the database. Adequate applications for classification include medical diagnosis, credit risk assessment, fraud detection and target marketing [Han et al., 2001].

**2.5.2.2 Association analysis**

Association rules show dependency relationships between rows and attributes in a database. Mining association rules is particularly important when trying to find relevant associations among items in a given customer transaction. These are typically expressed in the form X→Y (C, S) where X, Y are disjoint sets of database attributes, C is the confidence or condition probability $P(Y|X)$ and S is support or the probability $P(Y \cup X)$ [Hipp et al., 2000].

**2.5.2.3 Clustering**

Clustering algorithms also called unsupervised classification, is the process of grouping physical or abstract objects into classes of similar objects. Clustering analysis helps to construct meaningful partitions of a large set of objects based on a divide methodology, which decomposes a large-scale system into smaller components to simplify design and implementation this system [Sousa et al., 1998]. Unlike the concept of classification techniques, clusters are not pre-specified but rather emerge from the inherent similarity and dissimilarity among objects [Han et al., 2001].

**2.5.2.4 Outlier detection**

The objective of outlier detection technique is to discover set data objects those are abnormal, anomalies or unpredicted with the general behaviour of others data objects. These objects may be considered inconsistent, noise, and error with the remainder of data objects. As the result, many data mining techniques try to ignore, eliminate or minimize the influence of them [Knorr et al., 1997]. Other techniques consider the analysis of these objects may carry interested, unexpected, and implicit knowledge. Outlier detection can be beneficially used in many applications such as credit card fraud, voting irregularity and server weather prediction [Han et al., 2001] and [Ng 2001].

## 2.5 Conclusion

KDD process offers a great promise in helping organizations to discover hidden knowledge in their data. KDD process need to be guided by experienced users who understand the business, the data, and the general nature of the analytical methods involved in the organization. The main goal of KDD process is helping in decision making process.

In this chapter, a survey on KDD process was introduced and its importance in many applications. In addition, a survey on KDD's steps is presented especially data warehousing and data mining. Data mining is considered the main step in KDD process to find patterns and relationships between these patterns by building models. Data mining has many techniques divided into four main categories classification, association rules, clustering, and outlier detection.

In this thesis, we will focus on the techniques of KDD process that can be used in spatial databases especially outlier detection algorithm. The spatial KDD process will be presented in the next chapter.

# Chapter 3

# Spatial knowledge discovery

**3.1 Introduction**

**3.2 Spatial Data Warehouse (SDW)**

**3.3 Spatial Data Mining (SDM)**

**3.4 Conclusion**

## 3.1 Introduction

Intelligent GIS is one of the promising topics which are appeared in spatial domain. Intelligent GIS aims at integrating GIS with other computer sciences such as ES, DW, DSS, or KDD to beneficially exploit large amount of data stored in spatial databases. Geographic Knowledge Discovery (GKD) tries to exploit the great success that have been achieved by data mining and knowledge discovery techniques in databases by extending knowledge discovery techniques to be appropriate to be applied into spatial databases .

GKD is defined as "A process of extracting information and knowledge from massive geo-referenced databases" [Miller 2003].The classical KDD techniques are not sufficient to be directly applied into spatial datasets, because spatial databases highly differ from traditional databases[Shekhar et al., 2003a].The main difference between spatial databases and classical databases is the nature of the spatial objects. The objects in spatial databases are dependent on each others. The weight of this dependence inverses proportionally with the distance among these objects according to the first low of geography which states that "Every thing is related to every thing else, but nearby things are more related than distance things"[Tobler, 1979]. whereas the traditional databases are built on the fact that data samples are independent, so the techniques and algorithms of traditional data mining should be altered to take into consideration this property which is called spatial autocorrelation [Han et al., 2001] and [Shekhar et al., 2003b].

Spatial autocorrelation is not the only difference between the spatial knowledge discovery and the traditional knowledge discovery process. There are other factors such as the complexity of the spatial object and relationships among them, the heterogeneous and ill-structured nature of spatial data [Miller 2003].

## 3.2 Spatial Data Warehouse (SDW)

The wide use of DW structures and OLAP tools, for performing analysis to non-spatial data, initiates the interest to extend DW to be applied into spatial domain. Nevertheless, the transition from a transaction-oriented to an analysis-oriented in spatial datasets is not obvious [Zimányi 2003]. Spatial data representation, storage, retrieval, queries, and displaying data are different from the non-spatial data, thus we can not directly use the gained experience from non-spatial data with spatial data [Shekhar et al., 2001].

### 3.2.1 Spatial data warehouse applications

Spatial data warehouses contain spatial data, such as satellite images, aerial photography and non-spatial data such as the number of populations and the road capacity. Examples of spatial data warehouses include the US census dataset [Ferguson 1999], earth observation system archives of satellite imagery [USGS 1998] and highway traffic measurement archives [Lu et al., 2002].

The design of SDBs has been done usually based on specific software features, while these tools are more concerned with the geometry and topology of the objects than the modeling according to specific application or analysis needs, and the modeling tools were appeared after the development of SDB systems and applications. As the result, there is no common accepted conceptual model in spatial domain. . Besides that, there is no a very well defined conceptual modeling approach in DWs. The previous factors give the confusion of using the SDW term. Thus there is no formal definition of a SDW and different authors propose it according to their application needs [Bédard et al., 2001].

The word SDW is used in SDBs or GISs mostly when[Zimányi 2003]:-

- Huge volume of spatial data exists [Magon et al., 2001]
- The spatial data integration or the problem of multiple representations has to be solved [Bédard et al., 2002].
- Managing the metadata for spatial data [Brodeur et al., 2000]
- General analysis purposes referring to spatial data are existed [Zimányi 2003].
- Supporting decision-making process based on map representations [Grey 2003].
- Spatial aggregations need to be performed [Berkel 2003].

### 3.2.2 Spatial data warehouse components

Data warehouse components in the spatial domain consist of two main sets; measures and dimensions.

### 3.2.2.1 Dimensions in spatial domain

Dimensions in SDW are classified in three types [Bédard et al. 2001]

1. **Non-geometric spatial dimension:** This is a dimension whose primitive level and all of its higher generalization levels are non-geometric data such as product dimension, as depicted in (Figure 3-1), could start with the names of products, and its generalization levels will also be non-geometric, such as brand and categories.

2. **Geometric-to-non-geometric dimension:** This is a dimension whose primitive level is geometric but whose generalization levels, starting at a certain higher level, becomes non-geometric such as a client dimension in (Figure 3-1) is represented by points, geometric data, this level is the finest granularity level in this spatial dimension but its generalizations would be the names of city and country, which are non-geometric.

3. **Fully geometric spatial dimension:** This is a dimension whose primitive levels and all of its higher generalization levels are geometric. For example, polygons of villages, which are geometric data, are generalized to polygons of cities which are also geometric. Client dimension in (Figure 3-1) may also be fully geometric if the higher generalization levels such as city, state, and country also represented by polygons in maps [Bédard et al. 2001].

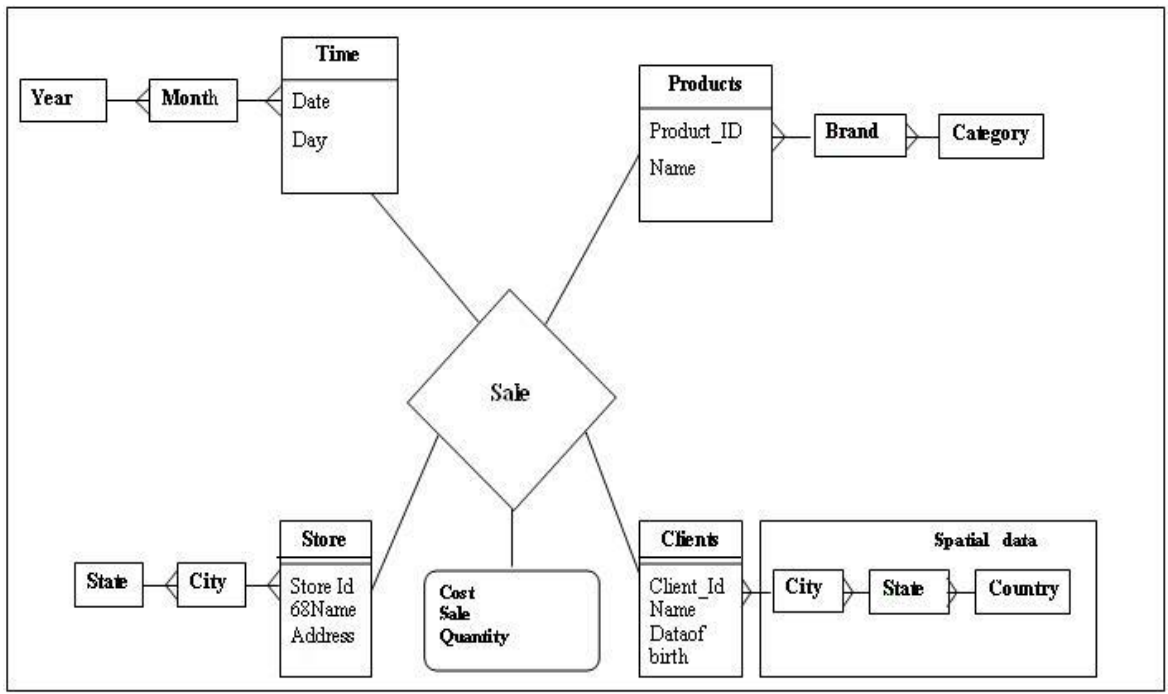

**3.** Figure 3-1: Star schema contains spatial dimension hierarchy

Measures within a SDW are classified into the following two types [Zimányi    [Zimányi 2003]

- **A numerical measure**; contains only numerical data such as sale in (Figure 3-1) is a numerical measure of a certain client and during the generalization may get the total sales of all clients.

- **A spatial measure;** contains a collection of pointers to spatial objects. For example, if the user is interested in analyzing the accidents' locations taking into account the different insurance types (coverage A, B, etc.), categories (cars, trucks, etc), and particular clients' data as depicted in (Figure 3-2), then the fact table will contain spatial measure, location, which represents the locations of the accidents. In coarse granularity, these pointers will point to the locations of accidents of a certain user , in a certain insurance type , and in a certain date, in higher level it will contain pointers to all locations of accident which are made by this user with all insurance categories in all times. [Zimányi 2003].

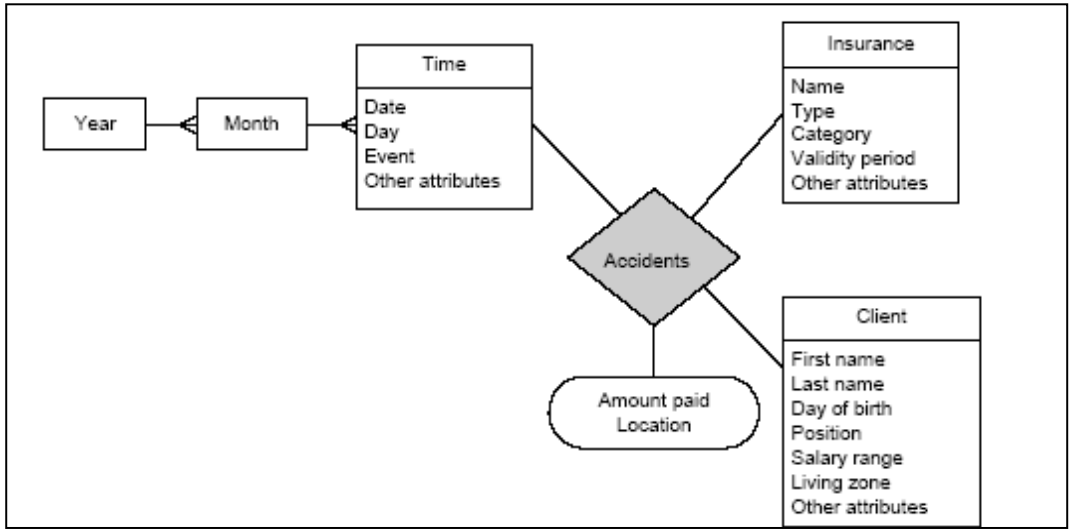

Figure 3-2: Star schema contains spatial measure

### 3.2.3 Map cube         [Zimányi 2003]

Map cube extends the concept of data cube to the spatial domain. A map cube is defined as an operator which takes the input parameters such as base maps, base tables and cartographic preferences to generate an album of maps for analysis. It is built from the requirements of a spatial data warehouse to aggregate data across many dimensions and to look for trends or unusual patterns related to spatial attributes. [Lu et al., 2002].

A map cube could be seen as a data cube with cartographic visualization of geometrical dimension to generate an album of related maps. A map cube adds more capabilities to traditional GIS such as roll-up, drill-down and slicing and dicing. It can benefit the processes of analysis and decision making which are based on spatial data warehouses [Bedard 1999]. Map cube inherits its ideas from three different domains namely data warehousing, Visualization and GIS as depicted in (Figure 3-3).

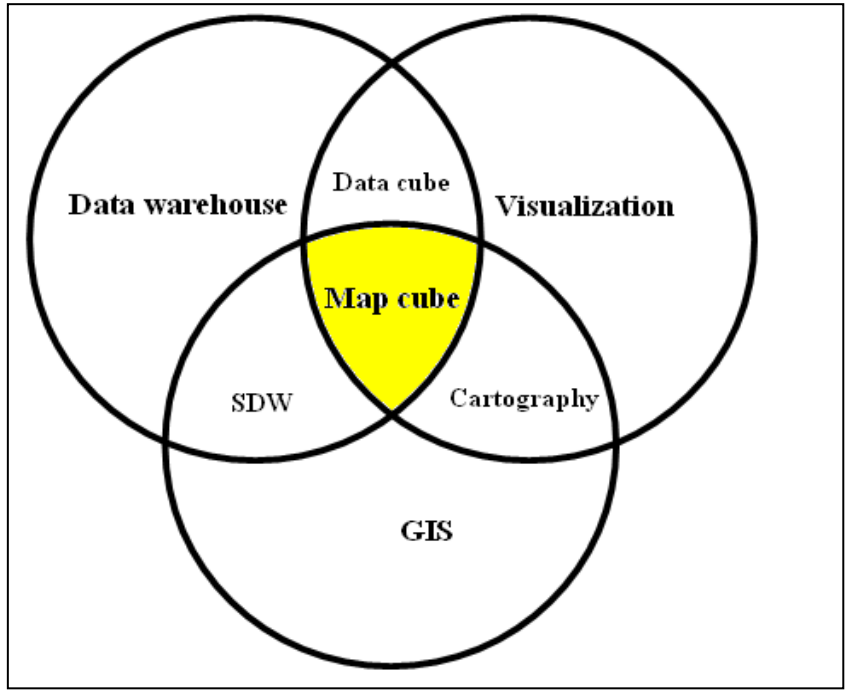

Figure 3-3: The relationship between map cube and three parent domains

[Lu et al., 2002]

A map cube consists of a lattice of cuboids; each of them represents a certain level of hierarchy. It uses a set of aggregate functions to compute statistics for a given set of values within each cuboid. These aggregate functions include sum, average, and centroid. The calculation of the aggregate functions may use other geometric operations such as geometric union as depicted in (Figure 3-4) [Lu et al., 2002].

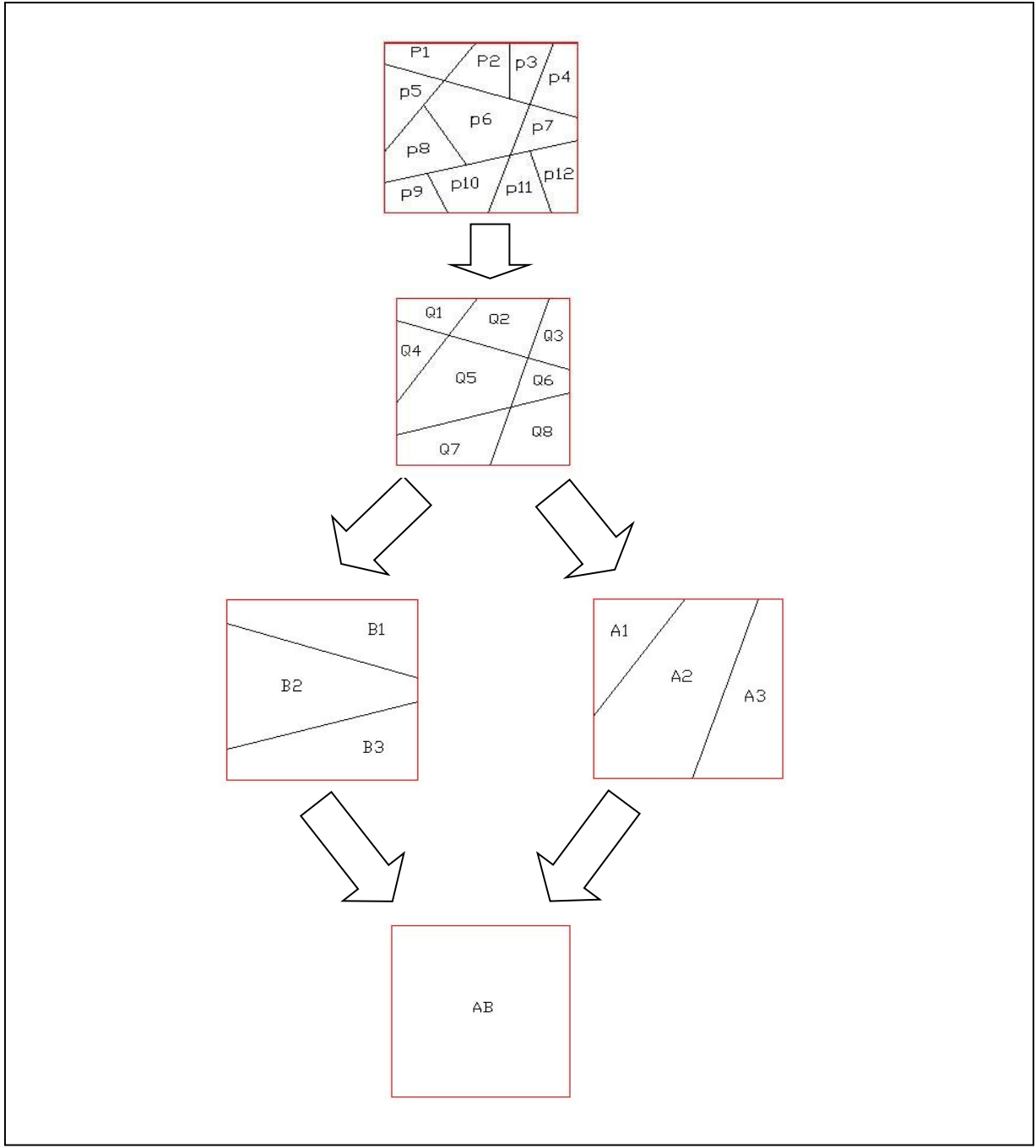

Figure 3-4: An example of geometric aggregation

[Lu et al., 2002]

## 3.3 Spatial Data Mining (SDM)

The main patterns types in traditional data mining; classification, association rules, clustering, and outlier detection have spatial extensions to be applied into spatial domain.

### 3.3.1 Spatial classification

Spatial classification techniques map spatial objects into meaningful classes but taking into consideration spatial factors such as distance, direction, accessibility, connections and cost [Miller 2003]. Spatial buffers are used to classify objects based on attribute similarity and distance-based proximity [Koperski et al 1998] [Ester et al. 1997] and [Ester et al, 2001]. One of the most popular techniques in Spatial Classification Auto-regression (SAM) which has been used in many applications with geo-spatial datasets [Kazar et al., 2005].

### 3.3.2 Spatial association

Spatial association rules are association rules that contain spatial predicates in their precedent or antecedent [Koperski et al., 1996]. A specific type of association rule is discovering a co-location pattern. These patterns are subsets of spatial objects that are usually located together [Huang et al., 2004] and [Xiong et al., 2004].

### 3.3.3 Spatial clustering

Spatial clustering algorithms exploit the topological relationships among spatial objects to determine inherent groupings of the input data [Miller 2003]. Clustering methods involve neither prior information nor number of clusters, so the cluster is called unsupervised learning. The cluster are formed based on the similarity criterion which is determined by relationships among spatial objects[Shekhar et al., 2003 b]

### 3.3.4 Spatial outlier detection

Spatial outlier detection techniques use spatial topological information to detect outlier objects. These objects have abnormal behavior with respect to its nearby objects [taha et al., 2004].

## 3.4 Conclusion

In this chapter, we presented a survey on geographic knowledge discovery process and its steps. The geographic knowledge discovery is an extension of KDD process into geographic databases. This new branch of KDD process tries to exploit large amount of data stored in spatial databases to produce smart knowledge to help in decision making process. The main difference geographic knowledge discovery process and traditional KDD process is spatial autocorrelation property which results from the difference between the nature of the spatial datasets and the traditional datasets since the objects in a spatial databases are dependent on other objects. The weight of this dependence is inversely proportional to the distance between these objects. Whereas the traditional databases are built on the fact that data samples are independently generated. Therefore, the techniques and algorithms of traditional KDD should be altered to take into consideration this property.

Spatial data mining is the main step of spatial knowledge discovery process. Spatial classification, spatial clustering, spatial association, and spatial outlier detection are the most important types of patterns in spatial data mining. In next chapter, different techniques of spatial outlier detection will be discussed in more details.

# Chapter 4

# Outliers detection algorithms in spatial domain

**4.1 Introduction to spatial outliers detection**

**4.2 Outliers detection algorithms in spatial domain**

**4.2 Homogeneous dimensions**

## 4.1 Introduction

Outliers are data objects which are abnormal, unpredicted or inconsistent with the other data objects in the domain. The behavior of these objects highly differs from the general behavior of other objects in the domain, so these objects may be considered as noise and error and may be ignored in the analysis. Recently, some data mining researches show the importance of analyzing the behavior of these objects [Miller and Han 2001].

Spatial outliers can be defined as a set of data objects that are displayed as abnormal, unexpected or inconsistent with neighboring objects which have an effect on the outliers objects through a predefined spatial neighborhood relationship [Miller and Han 2001] and [Miller 2003]. The spatial neighborhood relationship is based on a set of spatial factors such as distance, number of direct connections, cost and relative accessibility between objects… etc. Spatial outliers can be useful in many areas such as transportation, ecology, and location based services [Ester et al., 2001] and [Shekhar et al., 2002a].

## 4.2 Outliers detection algorithms in spatial domain

In outlier detection algorithms, we should distinguish between two main sets of attributes; one set of attributes is used for defining neighboring objects, and the other set is used for comparison of an object with its neighbors to detect spatial outliers. Classical researches in spatial data mining classify outliers in spatial datasets into two categories as depicted in (Figure 4 -1); one dimension and multi-dimension.

The first category, one dimension, researchers use only one of the non-spatial attributes for comparison purposes. In this case, the neighbor features are ignored which means ignoring spatial attributes and spatial relationships between objects. On the other hand, multi-dimension category uses some spatial or non-spatial attributes for defining neighbors and others for comparison purposes. Multi-dimension outlier detection techniques may be further divided into two subgroups, namely; homogeneous dimensions, and spatial outliers [Shekhar et al., 2003a].

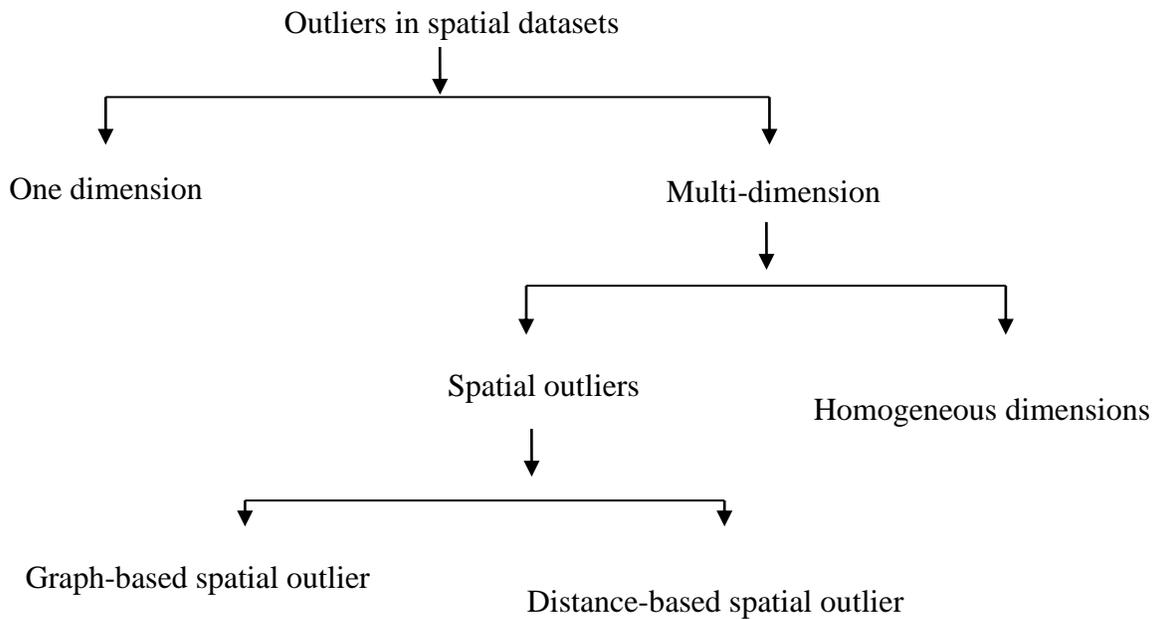

Figure 4-1   Classification of outlier detection techniques in spatial dataset

## 4.3 Homogeneous dimensions

Homogeneous dimensions techniques use spatial factors such as distance between two villages and non-spatial attributes such as number of populations in each village for defining neighbors and other spatial and non-spatial attributes for comparison purposes. There is no separation between spatial and non-spatial attributes. Many outlier detection methods were recently proposed in homogeneous dimensions outliers taking into consideration distance as a measure of outliers [Breunig et al., 1999], [Knorr et al., 1997], [Knorr et al., 1998], [Knorr et al., 2000]and [Ramaswamy et al, 2000]. In [Knorr et al., 1997], [Knorr et al., 1998], and [Knorr et al., 2000], the term distance-based outliers is defined as an object O in dataset T to be distance-based outlier DB (p,d)-outlier, where p is a fraction of all objects in T that lie within a distance greater than d from this object.

In [Ramaswamy et al, 2000] the authors find the K-nearest object for each object in a dataset, and then they rank all objects descendingly by the K-nearest neighbor distance, and then they define the top N objects as outliers. But in [Breunig et al., 1999] the data is stored in clusters. In each cluster, outlier detections tests have been done independently from other clusters. The resulting outliers are called local outliers because they are relative to each cluster.

These methods have many disadvantages namely, they do not distinguish between spatial and non-spatial attributes, and they detect global outliers not spatial outliers. Global outliers are objects that have a high difference in their attributes from the normal of non-spatial attributes of other objects in the space, whereas spatial outliers mean objects that have a high difference in their non-spatial attributes from the normal of non-spatial attributes of their spatial neighbors [Shekhar et al., 2003a].

## 4.4 Spatial Outliers

Spatial outlier algorithms also use both spatial attributes and non-spatial attributes, but they use spatial attributes for defining neighboring objects and use non-spatial attributes for comparison purposes [Shekhar et al., 2003b]. Spatial outlier detection techniques can be further divided into two subcategories; graph-based spatial outliers and distance-based spatial outliers. The difference between them lies in the definition of spatial neighborhood relationships [Shekhar et al., 2002a].

### 4.4.1 Graph-based spatial outliers

In the graph-based spatial outliers, the set of spatial neighboring objects to a certain objects contains all objects that have direct connections to this object [Shekhar et al., 2002a] and [Shekhar et al., 2001]. In the spatial data set which is shown in (Figure 4-2), then the set of graph-based neighboring objects, to object A, will include all objects that are directly connected to A {B, D, E} as shown in (Figure 4-3).

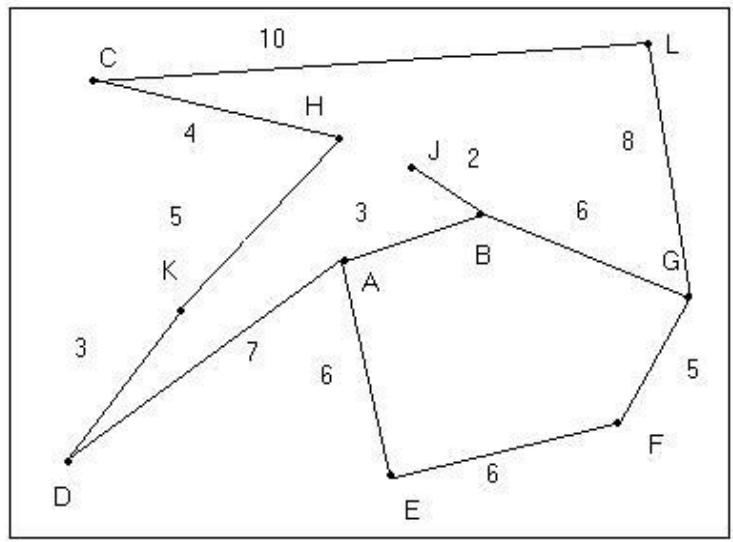

Figure 4- 2  spatial data set

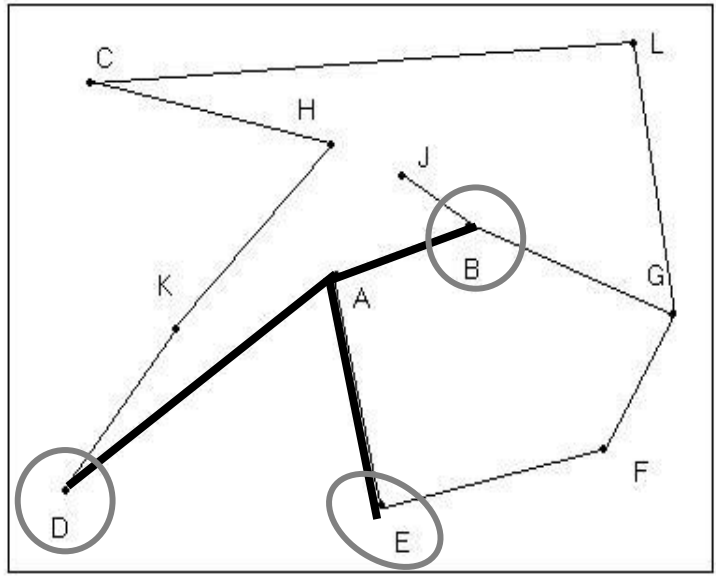

Figure 4- 3 Graph-based neighbors of node A

### 4.4.2 Distance-based spatial outliers

Distance-based spatial outlier techniques calculate spatial neighborhood relationships based on between objects by making a buffer zone around each object to discover its neighbors [Shekhar et al., 2003a]. For example, the spatial data set which is shown in (Figure 4-2), then the set of distance-based neighboring objects, to object A, will include all objects that lies inside the buffer zone of A {B, J,H,K} as depicted in (Figure 4-4).

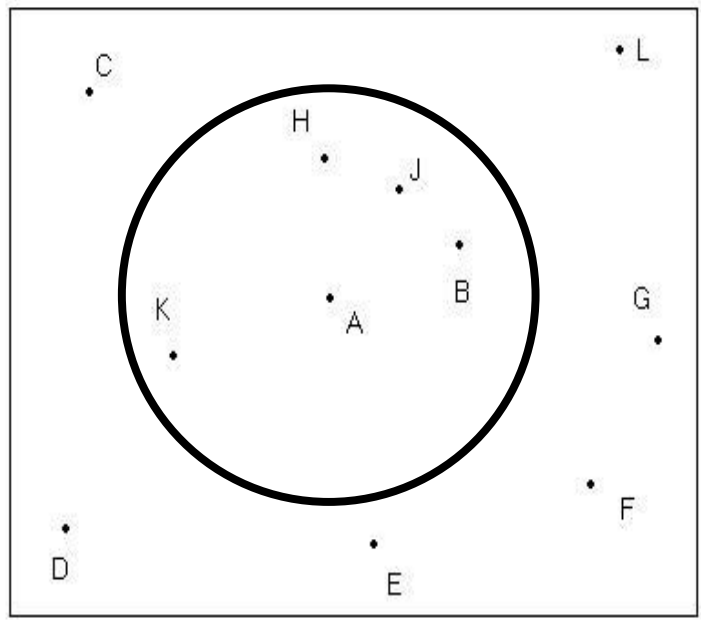

Figure 4- 4 Distance-based neighbors of node A

## 4.5 Conclusion

Spatial outlier detection is one of the main types of patterns in spatial data mining. It aims at finding objects which have extreme abnormal behaviors with respect to the behaviors of their spatial neighbors. In this chapter, a different types of outlier detection algorithms in spatial domain were presented. Graph-based spatial outliers and distance-based spatial outliers are the most popular types of spatial outliers. The difference between them lies in the definition of spatial neighborhood relationships.

Although a large amount of algorithms try to discover spatial outliers, they have many limitations.

First, most of spatial outlier algorithms assume that all neighbors to a certain object have the same weight of effect on this object, but the nearest object should have the largest weight of effect according to the first law of geography. Second, there is no standard to define the spatial neighborhood relationships; some algorithms leave these relationships to be defined manually by users. Other algorithms use external software to define them. This leads to different results for the same application using the same factors. Third, classical researches separate between graph-based spatial neighbors and distance-based spatial neighbors. But in many realistic applications, we should use connections and distances to discover spatial outliers. Last, such algorithms are designed to be applied only to points and no extensions of these algorithms could be applied to more complex objects such as lines or polygons.

# Chapter 5

# A proposed model for spatial outlier detection

**5.1 Limitation of classical spatial outlier detection algorithms**

**5.2 The proposed model**

**5.3 Spatial outlier Definition Adaptation**

## 5.1 Limitations of classical spatial outlier detection algorithms

Although large number of classical techniques that aim at discovering spatial outliers, but these techniques still have many limitations such as:

1. Most of spatial outlier algorithms assume that all neighbors to a certain object have the same weight of effect on this object, but the nearest object should have the largest weight of effect according to the first law of geography [Tobler 1979].
2. There is no standard to define the spatial neighborhood relationships; some algorithms leave these relationships to be defined manually by users. Other algorithms use external software to define them. This leads to different results for the same application using the same factors.
3. Classical researches separate between graph-based spatial neighbors and distance-based spatial neighbors. But in many realistic applications such as transportation and health, we should use graph and distances neighbors to discover spatial outliers [Taha et al., 2004].
4. Classical techniques are designed to be applied only to points and there is no extensions of these algorithms could be applied to more complex objects such as lines or polygons.

## 5.2 The proposed model

We try to solve limitations of the classical spatial outlier detection techniques by proposing a new model. This new model is based on weighted spatial neighborhood relationship.

### 5.2.1 The proposed weighted spatial neighborhood relationship

The first limitation of classical techniques in spatial outlier detection algorithms is the assumption, which supposes that all neighboring objects have the same effect on a certain object, which is not realistic. In fact each neighbor object has its own effect on the specified object. Applying this concept on distance-based spatial outlier algorithms, the nearest object has the largest effect. For graph-based spatial outlier algorithms, the objects that have direct connections have larger effect than of that object which has indirect connection. Also, the object that has the smallest cost with respect to a specified object has the largest effect on it.

So we suggest to change the classical form of spatial neighborhood relationship from (X,Y) to the Weighted Spatial Neighborhood Relationship(WSNR) form (X, Y, weight$_{XY}$); which means that an object X is affected by an object Y, and weight$_{XY}$ is the quantity of effect. This weight of effect must be between 0 and 1. The sum of weights that affect on the same object must be equal to 1. We should distinguish between weight$_{XY}$ and weight$_{YX}$. From this parameter, weight of effect, the objects that have the largest and the smallest effects on a specified object could be defined.

### 5.2.2 Calculating the expected value of spatial object

After we change the form of spatial neighborhood relationship to the new form we should provide a method to calculate the weight of effect. In the classical model, they suppose the same weight of effect to all neighbors of a certain spatial objects, which means that they implicitly use normal average. If we have an object r, which has N neighboring objects, then the weight of effect will be calculated as depicted in equation (5-1).

$$E(r) = \sum F(i) / N \qquad \text{For } i=1,\ldots N \qquad (5\text{-}1)$$

Where $F(i)$, $E(r)$ are the actual attribute value at object i, and the expected attribute value at object r respectively, and N is the number of neighboring objects relative to object r.

According to our proposed model equation (5-1) is modified to equation (5-2) as follows:

$$E(r) = \sum W_{ri} F(i) \qquad \text{For } i=1,\ldots N \qquad (5\text{-}2)$$

Where $W_{ri}$ is the weight of the effect from point i on point r, and the summation of $W_{ri}$ must be 1.

### 5.2.3 Calculating weight of effect

The problem has been reduced to how to calculate this weight automatically. A simple method to calculate this weight is also proposed here through a single spatial factor, namely, the distance between the objects. An extension of this method to cover many spatial factors is presented, controlling the influence of each. Since the distance is inversely proportional to the weight of the effect, the longest distance between objects indicates the lowest effect between them as indicated by equation (5-3).

$$W_{ri} = (1/D_{ir}) / \sum( 1/D_{ir}) \qquad \text{For i=1, …. N} \qquad (5\text{-}3)$$

Where $D_{ir}$ is the distance between the two points i,r .

Considering the effect of $R_{ir}$, which means the number of direct connections between the two points i and r, the largest value of $R_{ir}$ indicates the easiest movement from i to r, which means that number of direct connections is directly proportional to the weight of the effect. The corresponding weight can be expressed as follows:

$$W_{ri} = ( R_{ir}) / \sum( R_{ir}) \qquad \text{For i=1,……N} \qquad (5\text{-}4)$$

Taking into consideration the previous two factors together, we have:

$$W_{ri} = (\alpha /D_{ir}) / \sum( 1/D_{ir}) + (1- \alpha) (R_{ir}) / \sum( R_{ir}) \qquad \text{For i=1,…. N} \qquad (5\text{-}5)$$

Where α is the coefficient of the effect of distance which must be 0< α <1.

Additionally, the cost factor to move from point i to point r, $C_{ir}$, is inversely proportional to the weight. Thus, the overall weight can be formulated as:

$$W_{ri} = (\alpha /D_{ir})/\sum(1/D_{ir}) + (\beta R_{ir}) / \sum( R_{ir}) + (\delta/C_{ir} ) /\sum( 1/C_{ir})$$
$$\text{For i=1,….. N} \qquad (5\text{-}6)$$

Where α, β and δ are coefficients to distance, number of direct connections and cost, respectively, where α + β + δ = 1.

Equation (5-6) merges the most common three factors in point data types, which are distance, number of direct connections and minimal cost.

### 5.2.4 Methods for calculating spatial factors automatically

We also provide methods to calculate these factors automatically. These methods are:

I. Distance, by making a buffer zone around each object with a radius r defined by the user, which means neglecting the influence of the distance of objects outside this area. All objects inside this area are considered neighbors to this object.

II. Number of direct connections, can be calculated by finding the number of direct links passing through this object.

III. Minimal cost between two points, by defining the number of direct and indirect connections between the two points, and then finding the minimal cost for these connections. We can also define a limit L for the cost factor. If the minimal cost is greater than L then we will ignore the relationship to this node.

### 5.2.5 Merging distance-based and graph-based spatial outliers

the separation between distance-based spatial outliers and graph-based spatial outliers, which is found in classical models, could be eliminated by applying proposed model. We also can calculate distance-based only by putting α = 1 and β = δ = 0, and if we want to ignore distance effect, then put β =1.

## 5.3 Spatial outlier definition adaptation

After the suggested changes have been made through the proposed model, the definitions of the spatial outliers and the family of algorithms provided [Shekhar et al., 2003], should be accommodated to these changes.

Consider spatial framework SF = (S,NB), where S is a set of locations {s1,s2,s3,…,s$_n$} and NB (S,S,W) where W is the weight of the effect and W ϵ ]0,1[, NB is the neighborhood relation over S. We define neighbors of the location x in S using NB as N(x), specifically N(x) = {y | y ϵ S, NB(x,y,W$_{yx}$), and 1>W$_{yx}$> 0 }.

**Definition:** An object O is an S-outlier (f, f$^N_{aggr}$, F$_{diff}$, ST) if ST {F$_{diff}$ [f(x) , f $^N_{aggr}$(f(x),N(x))]} is true, where f : S ⟶ R is an attribute function, f $^N_{aggr}$ : R$^N$ ⟶R is a weighted aggregate function for the values of f over neighborhood, R is a set of real numbers, F$_{diff}$ R x R  R is difference function , and ST : R { true , false }is statistical test procedure for determining statistical significance .

We can define the function as the non-spatial attribute f $^N_{aggr}$ = E $_{yϵN(x)}$ (f(Y)) as the weighted average of attribute values over neighborhood N(x), F$_{diff}$ (x) is called the difference function S(x) = f(x)- f $^N_{aggr}$ is the arithmetic difference between attribute function and neighborhood aggregate function f $^N_{aggr}$ .Let  μ$_{S(x)}$ ნ$_{S(x)}$ be the mean and standard deviation of the difference function F$_{diff}$, respectively. Then the significance test function ST can be defined as Z $_{S(x)}$ = | S(x) - μ$_{S(x)}$ / ნ $_{S(x)}$ | > Ө, where Ө is the residual error. The choice of Ө specifies the confidence level. For example in normal distribution, if 95% confidence level is required, Ө = 2.

## 5.4 Conclusion

In this chapter, we change the classical form of spatial neighborhood relationship from to the WSNR form. The new form based on giving each neighbor its own weight of effect. A model for calculating weight of effect based on spatial factors such as distance, cost, and number of direct connection is also presented.

The new model can eliminate the separation between distance-based spatial outliers and graph-based spatial outliers, which is found in classical models. This model enriches the spatial neighborhood relationship by using many spatial factors and controlling the weight of each spatial factor.

We also change the classical spatial outlier detection model to accommodate the changes on spatial neighborhood relationship. In the next chapter, experimental results prove that the new model decreases the error that results from difference between calculated and the actual value on average.

# Chapter 6

# Experimental results

## 6.1 An illustrative application domain

Egyptian Center for Women's Rights (ECWR) in participation of Social Research Center (SRC) at American University in Cairo (AUC) started a project for literacy that aims at decreasing the percent of illiteracy in the Egyptian villages especially among females. The project focuses on percent of illiterate females between 14 and 35 years old. They started with El-Fayoum governorate which contains 167 villages as depicted in (Figure 6-1). These villages are grouped around 6 cities (El Fayoum, Tamyah, Ebshoway, Senorus, Atsa and Youseef Elsdeek as shown in (Figure 6-2). For each village, we have available the percentage of illiterates among females, males, and total population as shown in (Table 5-1). In our application, we are interested in discovering outlier villages which have the percent value of illiteracy in females highly differs from the average percent value of their neighboring villages. This application is developed using (MapObjects 2.1), (Visual Basic 6), and (MS Access 2000.)

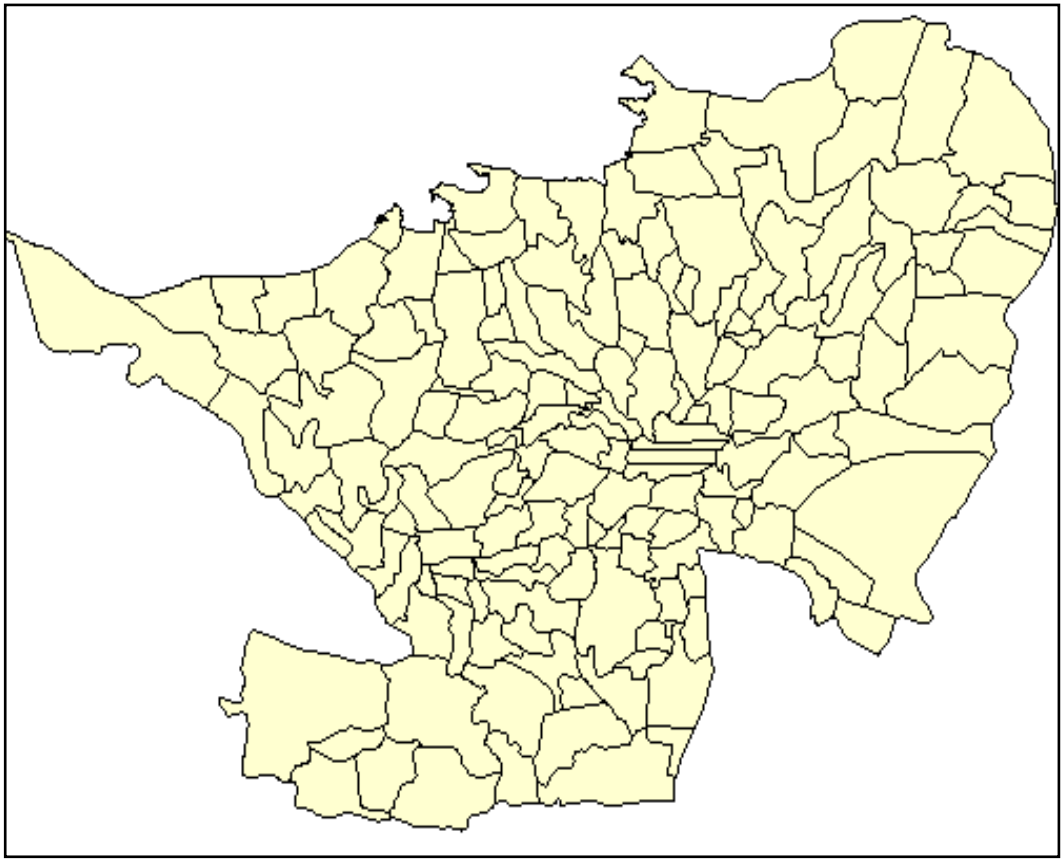

Figure 6-1: Map of El-Fayoum villages

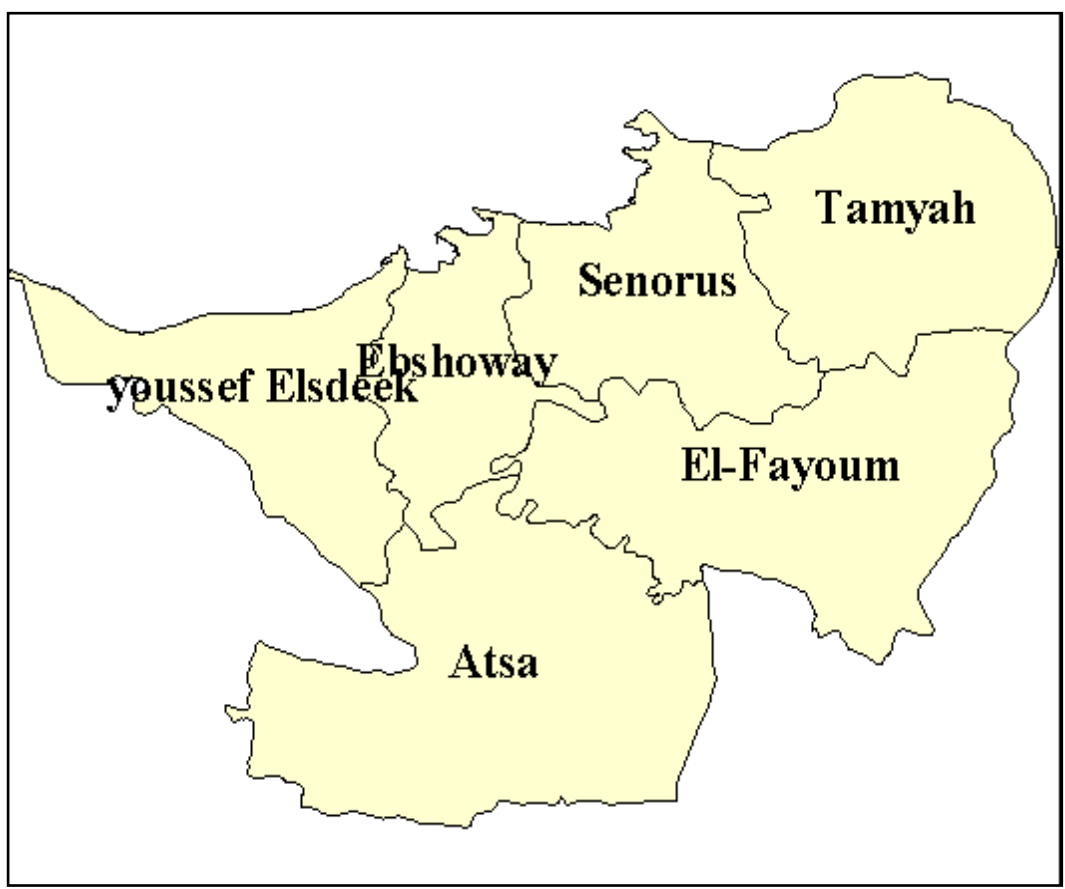

Figure 6-2: Map of El-Fayoum cities

| Markaz_ID | Village_ID | Village | Mail | Femail | total |
|---|---|---|---|---|---|
| 1 | 1 | أبجيج | 0.5026608612 | 0.664791901012 | 0.577633289986996 |
| 1 | 2 | أبو السعود | 0.2683438155 | 0.50234375 | 0.378827001106603 |
| 1 | 3 | الأعلام | 0.2551020408 | 0.419867549669 | 0.328884934756821 |
| 1 | 4 | البسيونية | 0.2752497225 | 0.590099009901 | 0.419053361471209 |
| 1 | 5 | الحادقة | 0.2918918919 | 0.502729257642 | 0.39177657098526 |
| 1 | 6 | الحميدية | 0.4396442186 | 0.599706744868 | 0.513955071477195 |
| 1 | 7 | السبخ فضل | 0.4106167057 | 0.710118505014 | 0.548780487804878 |
| 1 | 8 | الصالحية | 0.3971311475 | 0.737113402062 | 0.555749890686489 |
| 1 | 9 | العامرية | 0.320258949 | 0.479166666667 | 0.393691110200737 |
| 1 | 10 | العدوة | 0.249559305 | 0.509433962264 | 0.368262653898769 |

Table 6-1: Snap shot of illiteracy percentage in El Fayoum villages table

## 6.2 One dimensional outlier detection method

One dimension outlier detection method uses non-spatial attributes only for detecting outliers, which means ignoring spatial attributes and spatial relationships between objects. In this method, the expected value of illiterate females will be the same for each village and could be calculated from equation (6-1)

E( percent of illiterates ) = ∑ F (percent of illiterates) / N          For i=1,…. 167

$$= 58 \% \qquad (6\text{-}1)$$

If 95% confidence level is required, then $\Theta = 2$, then the outlier villages will have -2 > Fdiff  Or  Fdiff > 2  [Shekhar et al., 2003a]. The results of this test are shown in (Table 6-2).

| Markaz_ID | Village_ID | Village | percent of illitrate | Fdiff |
|---|---|---|---|---|
| 1 | 26 | قسم أول | 0.08 | -3.392 |
| 1 | 29 | قسم رابع | 0.14 | -2.993 |
| 1 | 28 | قسم ثان | 0.15 | -2.926 |
| 1 | 17 | دار الرماد | 0.18 | -2.727 |
| 1 | 27 | قسم ثالث | 0.26 | -2.195 |
| 4 | 325 | منشأة عطيفة | 0.27 | -2.128 |
| 3 | 216 | بندر إطسا | 0.27 | -2.128 |
| 3 | 238 | منشأة حلفا | 0.27 | -2.128 |

Table 6-2   Samples of results of Significance Test (ST) function of one dimensional outliers method

The same answer was got from the graphical test, scatter plot, as depicted in (Figure 6-3). We find that the acceptable difference is 0.3 then the outlier villages are that have the illiterate percentage in females out of the range [0.58-0.3, 0.58+0.3]. In other words, If a village has If (illiterate percentage < 28% OR illiterate percentage > 88%) then this village is consider outlier.

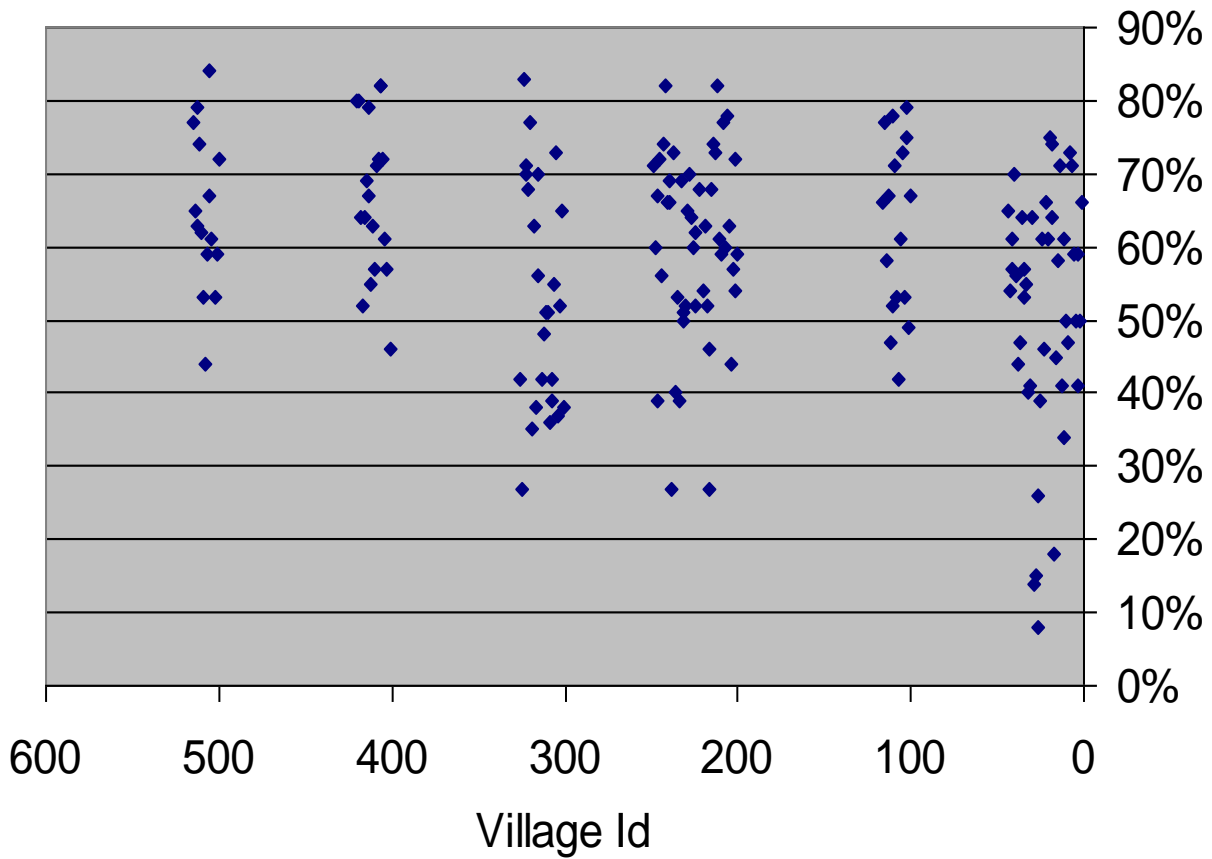

Figure 6-3: Results of graphical test of one dimensional outlier

## 6.3 Classical spatial outlier model

Classical spatial outlier model uses spatial attributes for defining neighboring objects and non-spatial attributes for detecting outliers. The classical spatial outlier model in this application has two limitations. First, the data type of objects is polygons not points. Second, we need to define neighboring objects. In this application, we define the spatial neighbors to a certain object as all adjacent objects which have a common line with that object, so the expected value of illiterate percentage in any village will be the average of illiterate percentages in its adjacent villages [Shekhar et al., 2003a] and [Taha et al., 2004].

If 95% confidence level is required, then $\Theta = 2$, then the outlier villages will have ($-2 >$ Fdiff Or Fdiff $> 2$). The results are shown in (Table 6-3) and (Figure 6-4).

| Markaz_ID | Village_ID | Village | percent of illitrate femails | Classical_Diff |
|---|---|---|---|---|
| 1 | 30 | كفور النيل | 0.64 | 2.50719435912913 |
| 4 | 317 | فيدمين | 0.38 | -2.10837522956018 |
| 1 | 26 | قسم أول | 0.08 | -2.2220392123951 |
| 1 | 29 | قسم رابع | 0.14 | -2.33001999608827 |
| 1 | 28 | قسم ثان | 0.15 | -2.41739918289262 |
| 3 | 238 | منشأة حلفا | 0.27 | -2.50051597034065 |
| 3 | 216 | بندر إطسا | 0.27 | -2.60565515446294 |
| 1 | 17 | دار الرماد | 0.18 | -2.60768486844215 |

Table 6-3: Samples of results of Significance Test (ST) function of classical spatial outliers model

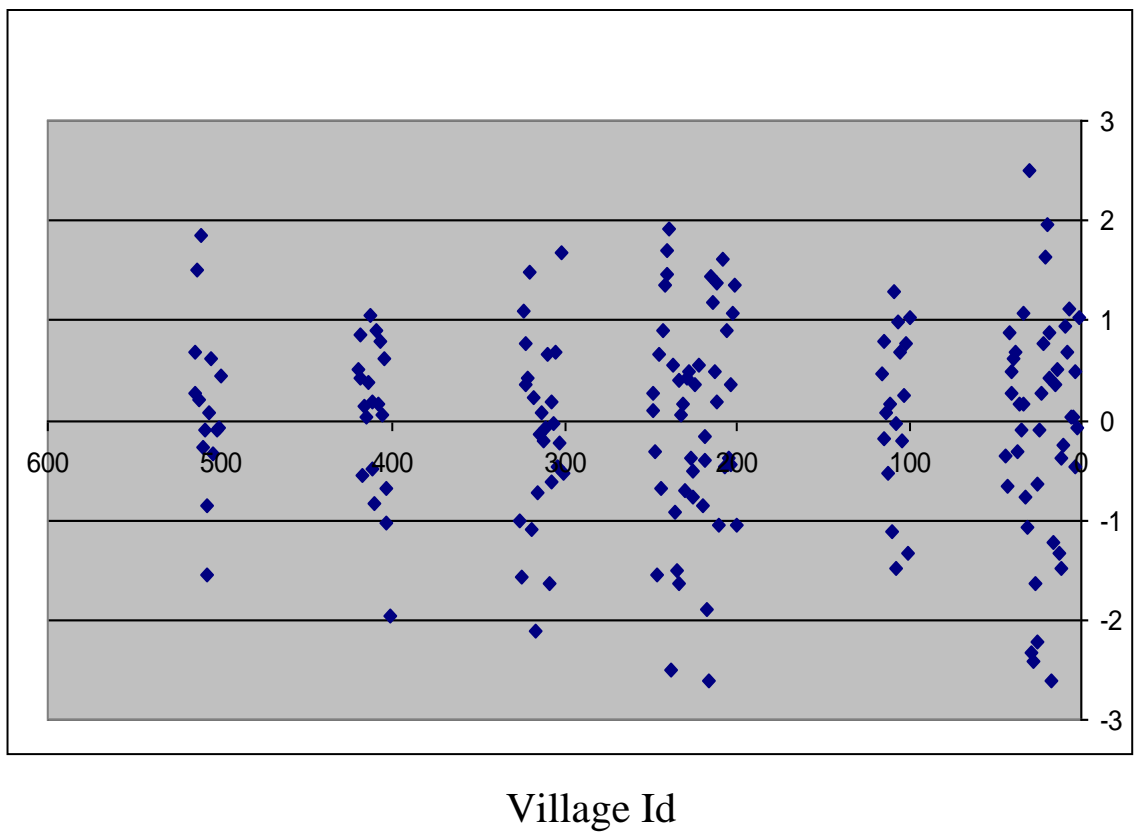

Village Id

Figure 6-4: Results of graphical test of classical model

## 6.4 The proposed model

To apply the proposed model in this application, some limitations have been considered, the same limitations which face us in classical model beside two new limitations. First, what are the spatial factors that are taken into consideration to calculate of the weight effect between neighboring objects? Second, how will we calculate the distance between two polygons?

We make the next assumptions to help in solving the problems:
- We use distance between objects and area of neighboring objects to calculate the weighted effect, the distance being inversely proportional to the weighted effect, while the area being directly proportional to the weighted effect.
- We consider the distance between objects as the distance between their centers.

If 95% confidence level is required, so we take $\Theta = 2$ so the spatial outlier object has $| S(x) - \mu S_{(x)} / \sigma_{S(x)} | > 2$. Samples of results of significance test function ST of our proposed model and the classical model are shown in (Table 6-4) and (Figure 6-5).

| المركز | رقم القرية | القرية | Percent of illiterate | Proposed_Fdiff |
|---|---|---|---|---|
| 1 | 26 | قسم أول | 0.08 | -2.23356990722518 |
| 1 | 17 | دار الرماد | 0.18 | -2.78113261661175 |
| 1 | 30 | كفور النخل | 0.64 | 2.45943452129421 |
| 3 | 216 | بندر إطسا | 0.27 | -2.74275657908981 |
| 3 | 217 | نطون | 0.46 | -2.00176256348505 |
| 3 | 238 | منشأة حلفا | 0.27 | -2.43846220240124 |
| 3 | 239 | منشأة ربيع | 0.69 | 2.08583349745509 |
| 4 | 317 | فدمين | 0.38 | -2.26780238294499 |

Table 6-4 Samples of results of Significance Test (ST) function of proposed spatial outliers Model

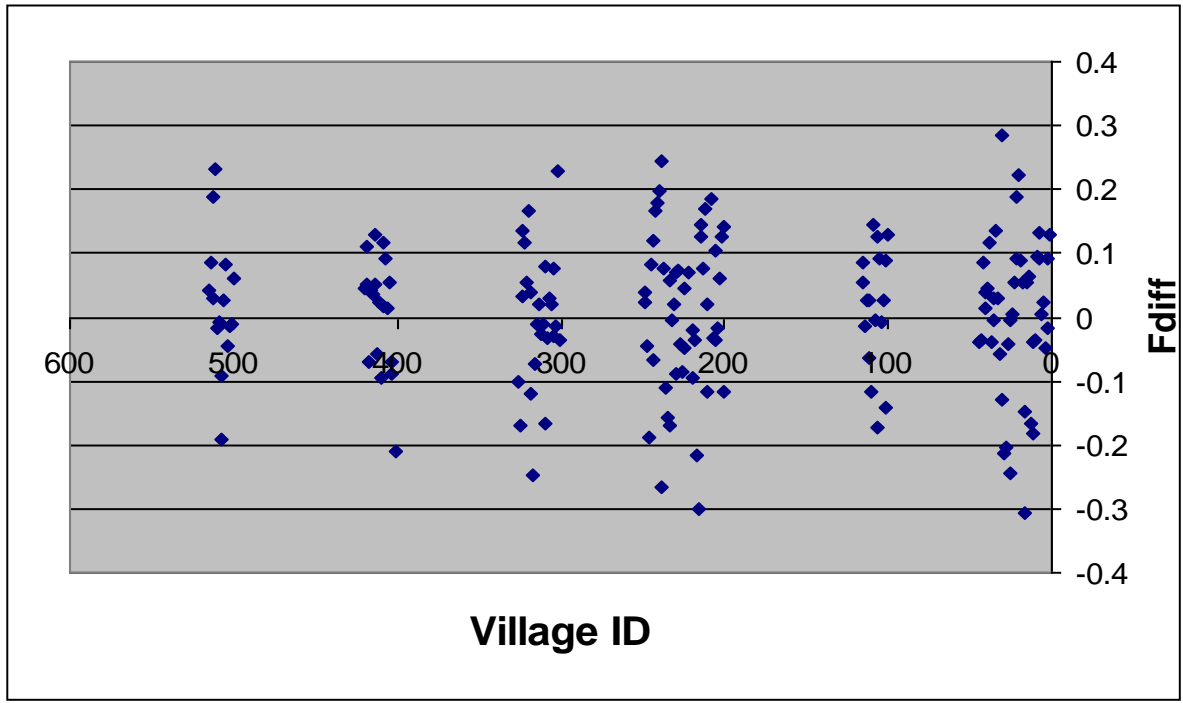

Figure 6-5: Results of graphical test of proposed spatial outliers' Model

## 6.5 A comparison between the proposed model and classical models

In this section, a compare will be held among the results of our proposed model [Taha et al., 2004] and those of applying classical spatial outlier model and one dimensional outlier[Shekhar et al., 2003 a].

**6.5.1 Classical spatial outlier model VS the proposed model**

In the proposed model, The village_Id 27 has 7 neighboring villages as depicted in (Figure 6-6) and the actual percent of illiterate females is 26%. The percent of illiterate females and different weight of effect values of the neighboring villages to this village, village_Id 27, are shown in (Table 6-5). We found that village_Id 29 is the nearest village, so it has the largest value (nearly 41% of the total effect). Whereas village_Id 42 has the smallest value (about 5%). The classical model assumes that the two villages (village_Id 29 and village_Id 42) have the same weight of effect (about 14,3%). This difference made our expected percent of illiterate females in this village 28%, while it is 45% in the classical spatial outlier model.

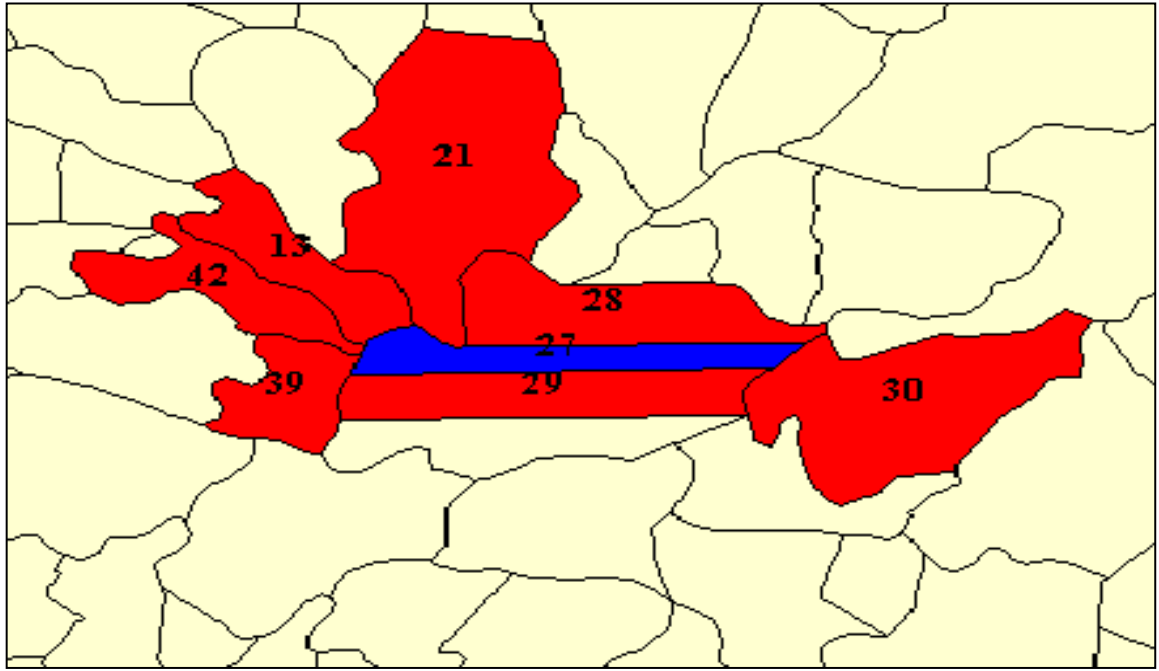

Figure 6-6  Village_Id 27 and its neighboring villages

| Nighbore Village Id | Weight Of Effect | Percent Of Illuterates Females |
|---|---|---|
| 29 | 41% | 14% |
| 28 | 26% | 15% |
| 21 | 8% | 61% |
| 39 | 7% | 56% |
| 13 | 7% | 41% |
| 30 | 6% | 64% |
| 42 | 5% | 61% |

Table 6-5 Table of weight of effect and percent of illiterate females for each neighboring village to village_id  27, ordered according to its effect

The difference error S(x) between our proposed model and the classical model can be calculated by equations (6-1) and (6-2)

S(x): 28% - 26% = 2%　　　　　for our proposed model　　　　(6-2)
S(x): 45% - 26% = 19%　　　　　for the classical model　　　　(6-3)

The difference between the square of difference error of the classical model and the proposed model is:

$(0.19)^2 - (0.02)^2 = 0.0357$ (6-4)

The percentage of improvement which is made by the proposed model with respect to one classical spatial outlier model in this village is:

$(0.19)^2 - (0.02)^2 / (0.19)^2 = 98.89\%$ (6-5)

We find the average percentage of improvement which is made by the proposed model with respect to classical spatial outlier model in this application is about 8%.

### 6.5.2 One dimensional outlier model VS the proposed model

We will also take about the same village, village-id 27, as an example. In one dimensional, the expected percent of illiterate females in all villages will be 58%, so expected percent of illiterate females in village-id 27 is 58%. But in our proposed model, the expected percent of illiterate females in village-id 27 is 28%. While the actual percent of illiterate females in village_id 27 is 28%.

The difference error S(x) between our proposed model and the one dimensional model can be calculated by equations (6-6) and (6-7)

S(x): 28% - 26% = 2%　　　　　for our proposed model　　　　(6-6)
S(x): 58% - 26% = 22%　　　　　for the classical model　　　　(6-7)

The difference between the square of difference error of the classical model and the proposed model is:

$$(0.22)^2 - (0.02)^2 = 0.048 \qquad (6\text{-}8)$$

The percentage of improvement which is made by the proposed model with respect to one dimensional model in this village is:

$$(0.19)^2 - (0.02)^2 / (0.19)^2 = 99.1\% \qquad (6\text{-}9)$$

We find the average percentage of improvement which is made by the proposed model with respect to one dimensional model in this application is about 42,4%.

# Conclusion and future work

**Conclusion**

**Future work**

# Conclusion

Geographic knowledge discovery is one of the promising topics in GIS which extends the KDD process to be applied spatial databases. GKD tries to exploit large amount of data stored in geographic databases to produce smart knowledge to help in decision making process. The main difference GKD process and traditional KDD process is spatial autocorrelation property. This property results from the difference between the nature of spatial dataset and traditional databases since the objects in a spatial databases are dependent on other objects. The weight of dependence is inversely proportional to the distance between these objects according the first geography law. Whereas the traditional database builds on the fact that data samples are independently generated. Therefore, the techniques and algorithms of traditional KDD should be altered to take into consideration this property.

Spatial outliers are the set of spatial objects which have extreme abnormal behaviors with respect to the behaviors of their spatial neighbors. Such neighboring have an effect on the outliers objects through a predefined spatial neighborhood relationship. There are two types of spatial outliers; graph-based spatial outliers and distance-based spatial outliers.

Although many algorithms try to discover spatial outliers, they have many limitations; first, most of spatial outlier algorithms assume that all neighbors to a certain object have the same weight of effect on this object, but the nearest object should have the largest weight of effect according to the first law of geography. Second, there is no standard to define the spatial neighborhood relationships; some algorithms leave these relationships to be defined manually by users. Other algorithms use external software to define them. This leads to different results for the same application using the same factors. Third, classical researches separate between graph-based spatial neighbors and distance-based spatial neighbors. But in many realistic applications, we should use connections and distances to discover spatial outliers. Finally, such algorithms are designed to be applied only to points and no extensions of these algorithms could be applied to more complex objects such as lines or polygons.

In this thesis, we proposed a new definition to spatial neighborhood relationship adding a new parameter which is called weight of effect; we also change the definition of spatial outliers to accommodate with the new definition of spatial neighborhood relationship. We analyze the current outlier detection methods in spatial data set, and then we merge between graph-based and distance-based spatial outliers.

The proposed is based on many spatial factors such as distance, numbers of direct connections, and cost. We extend the proposed model to be applied on more complex data types such as polygonal objects. In addition we provide experimental result from application of proposed model on an existing project for supporting literacy in Fayoum governorate in Arab Republic of Egypt (ARE). Experimental results prove that our model decreases the error that results from difference between calculated and the actual value on average 8%, and by 98% in some villages.

## Future work

- We still have many limitations in defining spatial neighborhood relationship in lines and polygons. In polygons we should take in consideration length of common arc, the shape of contour, and the cost to move from one polygon and other polygons. We also should redefine the model of detecting neighbors to allow not only adjacency polygons to be added to the set of neighboring objects. In lines, some spatial factors should be considered such as distance between two lines how to be measured, the intersection and location of intersection, the largest vertical and horizontal distance between two lines.
- The proposed model is applied using single non-spatial attribute in comparisons to detect outliers, if we use multiple non-spatial attributes in comparisons, many modification should be considered.
- The proposed model used to detect spatial outliers but it may help in discovering temporal and spatial-temporal outliers. We plan to investigate the definition of temporal and spatial-temporal neighborhood relationship.

- We plan to investigate the new neighborhood relation definition in other spatial data mining algorithms such as classification, clustering and co-location algorithms.
- Analyzing historical information of the spatial outliers can be beneficial in the DSS process, so we plan to make Spatial Outlier Data Warehouse( SODW).

# References

[ANL 2003] Alamos National Laboratory. "**Earth & Environmental Sciences. GISLab. Spatial Data Warehouse**"http://www.gislab.lanl.gov/data_warehouse.html, 2003.

[Bédard 1999] Bédard Y., "**Visual modeling of spatial databases towards spatial extensions and uml**". In Geomatica, 1999.

[Bédard et al., 2001] Bédard Y., Merrett T., and Han J. "**Fundamentals of Spatial Data Warehousing for Geographic Knowledge Discovery**". In Miller H. and Han J. (eds.) Geographic Data Mining and knowledge discovery. Taylor & Francis, 2001.

[Bédard et al., 2002] Bédard Y., Prouls M.J., Larrivée S., and Bernier E. "**Modelling Multiple Representations into Spatial Data Warehouses: a UML-based Approach**" In Symposium of Geospatial Theory, Processing and Applications, Symposium Ottawa, Canada, 2002.

[Berkel 2003] Berkel J. "**Data Warehouse; where to locate GIS**". http://gis.esri.com/library/userconf/proc97/proc97/to650/pap650/p650.htm, 2003.

[Breunig et al., 1999] Breunig,M.M, Kriegel,H.P., Ng,R.T., and sander,J., "**Optics-of: Identifying local outliers**" in proc.of PKDD 99 pargue, Czech Republic,1999.

[Brodeur et al., 2000] Brodeur, J., Bédard Y., and Proulx M.J "**Modelling Geospatial Application Databases using UML-based Repositories Aligned with International Standards**" in Geomatics. In Proc. of the ACM Symposium on Advances in Geographic Information Systems, ACM GIS, 2000.

[Burrough 1986] Burrough, P.A., " **Principles of Geographical Information Systems for Land Resources Assessment**". Clarendon, Oxford, 1986.

[Davis 1996] Davis E " **GIS: A Visual Approach**" Santa Fe: Onward Press, 1996.

[Eder et al.,2002] Eder J., Koncilia Ch., and Morzy T. "**The COMET Meta Model for Temporal Data Warehouses**" In Proc. of the 14th Int. Conf .on Advanced Information Systems Engineering, CAISE, 2002.

[ESRI 2003]  Environmental Systems Research Institute Http://www.esri.com.

[Ester et al., 1997] Ester, M., Kriegel, H.-P. and Sander, J. "**Spatial data mining: A database approach**" M. Scholl and A, Voisard (eds.) Advances in Spatial Databases , 1997.

[Ester et al., 2001] Ester,M., Kriegel,H.P.and Sander,J "**Algorithms and applications for spatial data mining**" (eds.) Geographic Data Mining and Knowledge Discovery, London: Taylor and Francis, 160-187,2001.

[Fayyad et al., 1996] Fayyad,U.M., Piatetsky-Shapiro,G., and Smyth.P., " **From data mining to knowledge discovery: An overview**" in (eds.) Advances in Knowledge Discovery and Data Mining, Cambridge, MA: MIT Press, 1996.

[Ferguson 1999] Ferguson, P., "**Census 2000 behinds the scenes**" In Intelligent Enterprise, October 1999.

[Gahegan et al., 2001] Gahegan,M., Wachowicz,M. Harrower,M. and Rhyne,T.M "**The integration of geographic visualization with knowledge discovery**" in databases and geocomputation," Cartography and Geographic Information Systems, 2001.

[Golfarelli et al., 1999] Golfarelli M., Rizzi F., "**Designing the Data Warehouse: Key Steps and Crucial Issues**" In Journal of Computer Science and Information Management, Vol. 2(3), 1999.

[Grey 2003] Grey B.," **Spatially Enabling an Incident Data Warehouse. GIS/Trans,Ltd**"     http://gis.esri.com/library/userconf/proc00/professional/ papers/PAP762/p762.html, 2003.

[Guting, 1994] Guting, R.H.,  "**An Introduction to Spatial Database Systems**" VLDB J., special issue on spatial database systems, vol. 3, no. 4 , 1994.

[Han et al., 2001] Han,J., and Kamber , M., " **Data Mining concepts and techniques** "Academic Press, ISBN 1-55860-489-8,2001

[Harvard 2005]   Harvard design school "**Vector GIS Procedures** "A Tutorial in GIS" http://www.gsd.harvard.edu/geo/manual/vector, 2005.

[Hipp et al., 2000] Hipp, J., Güntzer, U. and Nakhaeizadeh, G. "**Algorithms for association rule mining: A general survey and comparison**" SIGKDD Explorations, 2, 2000.

[Huang et al., 2004] Huang, Y. Shekhar, S. and Xiong, H. "**Discovering co-location patterns from spatial datasets: A general approach**" IEEE Transactions on Knowledge and Data Engineering (TKDE), 16(12), 2004.

[Hughes 1994] Hughes, D., "**An introduction to Geological Information Systems**" HMSO publication center ISBN 0 11 330612 1,1994.

[Inmon 1996] Inmon, W., "**Building the Data Warehouse**" John Wiley & Sons, 1996.

[Kazar et al., 2005] Kazar B. M., Shekhar S., Lilja D. J., Shires D., Rogers J., and Celik M., "**A Parallel Formulation of the Spatial Auto-Regression Model**" in the International Conference on Geographic Information GIS PLANET, Lisbon, Portugal, 2005.

[Keating et al., 1987] Keating, T., W. Phillips and K. Ingram, 1987. "**An integrated topologic database design for geographic information systems**" Photogrammetric Engineering and Remote Sensing Vol. 53. 1987.

[Kimball 96] Kimball, R., "**The Data Warehouse Toolkit**" John Wiley & Sons, 1996.

[Kimball et al., 98] Kimball R., Reeves L, Ross M., and Thornthwaite W. "**The Data Warehouse Lifecycle**". John Wiley & Sons, 1998.

[Knorr et al., 1997] Knorr,E., and Ng,R., "**A Unified Notation of Outliers: Properties and Computation**" in Proc. of the International conference on Knowledge Discovery and Data Mining ,1997.

[Knorr et al., 1998] Knorr,E., and Ng,R., "**Algorithms for distance-based outliers in large datasets**", in Proc. 24th VLDB conference, 1998.

[Knorr et al., 2000] Knorr,E., Ng,R., and Tucakov,V., "**Distance-based outliers: Algorithms and applications**" ,VLDB journal Vol 8, 2000.

[Koperski et al., 1996] Koperski,K., Adhikary,J., and Han,J., "**Spatial data mining: Progress and challenges**", In Workshop on Research Issues on Data Mining and Knowledge Discovery(DMKD'96), Montreal, Canada, 1996.

[Koperski et al., 1998] Koperski, K., Han, J., and Stefanovic N. "**An efficient two-step method for classification of spatial data**" Proceedings of the Spatial Data Handling Conference, Vancouver, Canada, 1998.

[Lakhan 1996] Lakhan V. C., "**Introductory Geographical Information Systems**" Chelsea Summit Press, 1996.

[Lu et al., 2002] Lu, C. T., Shekhar,S., Tan, X. , and Chawla, S., "**Map Cube: A Visualization Tool for Spatial Data Warehouses**" Proceedings of NSF workshop on Data Mining in GIS, 2002.

[Magon et al., 2001] Magon A., Aggarwal P. "**Data Warehouse: Different Alternatives**" In 22$^{nd}$ Asian Conference on Remote Sensing, 2001.

[Miller et al., 2001]  Miller, H. J. and Han, J  "**Geographic data mining and knowledge discovery: An overview**" in Miller, H. J., and Han, J. (eds.) Geographic Data Mining and Knowledge Discovery, London: Taylor and Francis, 2001

[Miller 2003]  Miller,H.J. "**Geographic Data Mining and Knowledge Discovery**" Handbook of Geographic Information, Science, in press, 2003

[NCGIA 1990] National Center for Geographic Information and Analysis, http://www.geog.ubc.ca/courses/klink/gis.notes/ncgia/toc.html,1990.

[Ng 2001]Ng, R. "**Detecting outliers from large datasets**" in H. J. Miller and J. Han (eds.) Geographic Data Mining and Knowledge Discovery, London: Taylor and Francis, 218-235, 2001.

[Parent et al., 1987] Parent, P. and Church R., "**Evolution of Geographical Information Systems as Decision Making Tools**" Proceedings, GIS ', pp. 63-71, ASPRS ACSM, Falls Church, VA, 1987.

[Pedersen 2000] Pedersen T.B. "**Aspect of Data Modeling and Query Processing for Complex Multidimensional Data**" PhD thesis. Aalborg University, Denmark, 2000.

[Qi et al., 2003] Qi, F. and Zhu, A.,X.,  "**Knowledge discovery from soil maps using inductive learning**," International Journal of Geographical Information Science, 17, 2003.

[Ramaswamy et al, 2000] Ramaswamy,S., Rastogi,R., and Shim,K., "**Efficient algorithms for mining outliers from large data sets**" in proceeding of the 2000 ACM SIGMOD International Conference on management of data Vol 29,ACM,2000.

[Shekhar et al., 1999] Shekhar,S., Chawla, S., Ravada,S., Fetterer, A., Liu,X., and Lu,C.T, " **Spatial Databases Accomplishments and Research Needs**" IEEE Transaction on Knowledge and Data Engineering, VOL. 11, NO.1, 1999.

[Shekhar et al., 2001] Shekhar, S., Lu C.T., and Zhang P., "**Detecting Graph-based Spatial Outliers: Algorithms and Applications**". Proceedings of the Seventh ACM SIGKDD International Conference on Knowledge Discovery and Data Mining, 2001, San Francisco, CA, 2001.

[Shekhar et al., 2002 a] Shekhar,S., Lu,C.T., and Zhang,P., "**Detecting graph-based spatial outliers**" in Intelligent Data Analysis 6,2002.

[Shekhar et al., 2002 b] Shekhar S., Zhang P., Lu C.T., and Liu R., "**Data Mining for Selective Visualization of Large Spatial Datasets**", 14th IEEE International Conference On Tools with Artificial Intelligence, 2002.

[Shekhar et al., 2003 a] Shekhar,S., Lu,C.T., and Zhang,P., "**A Unified Approch to Detecting Spatial Outliers**" GeoInformatica 7(2)- 139-166,2003

[Shekhar et al., 2003 b] Shekhar,S., and Chawla,S, "**A Tour of Spatial Databases**" Prentice Hall, ISBN 0-13-017480-7, 2003.

[Sousa et al., 1998]M. S. Sousa, M. L. Q. Mattoso & N. F. F. Ebecken," Data **mining: a database perspective**" Proc.Int.Conf.on PDPTA, SCREA Press, Las Vegas,1998.

[Star et al, 1990]Star, J.L. and J.E. Estes, 1990 "**Geographic Information Systems: An Introduction**", Prentice Hall, Englewood Cliffs, NJ. An introduction to GIS with a strong raster orientation, 1990.

[Taha et al., 2004] Taha A., Onsi H., and Hegazi O., and Nour Dein M., "**A model for spatial outlier detection** " Egyptian Informatics Magazine, Cairo University, Egypt, 2004.

[Theodoratos et al., 99] Theodoratos D., and Sellis T., " **Designing Data Warehouses**" In Data & Knowledge Engineering, Vol. 31, 1999.

[Tobler 1979] Tobler, W., "**Cellular Geography, Philosophy in Geography**" Dordrecht, Reidel: Gale and Olsson, Eds, 1979.

[Tomlinson 1987] Tomlinson, R.F., "**Current and potential uses of geographical information systems**" International Journal of Geographical Information Systems 1, 1987.

[USGS 1998] USGS "**National satellite land remote sensing data archive**" In URL http://edc.usgs.gov/programs/nslrsda/overview.html, 1998.

[Williams et al., 1998]Williams G., Hegland M., and Roberts S.," **A Data Mining Tutorial**" Second IASTED International Conference on Parallel and Distributed Computing and Networks (PDCN'98),1998.

[Xiong et al., 2004] Xiong,H., Shekhar,S., Huang,Y., Kumar,V., Ma,X., and Yoo,J.S. "**A Framework for Discovering Co-location Patterns in Data Sets with Extended Spatial Objects**" SIAM,2004 [Zimányi 2003] Zimányi E., "**Concepts and Methodolgical Framework for Spatio-Temporal Data Warehouse Design**" PhD thesis. University Libre of Bruxelles, Belgium 2003.